# Relating chemical bonding to physical properties: The origin of unexpected isotropic properties in layered materials

*Jiawei Zhang, Lirong Song, Mattia Sist, Kasper Tolborg, and Bo Brummerstedt Iversen*[*]

*Center for Materials Crystallography, Department of Chemistry and iNANO, Aarhus University, DK-8000 Aarhus, Denmark*
*E-mail:* bo@chem.au.dk

**Abstract**

Layered materials span a very broad range of solids ranging from van der Waals materials to highly complex crystal structures such as clays. They are commonly believed to have highly anisotropic properties, which is essentially attributed to weak interlayer interactions. The layered $Mg_3Sb_2$ structure is currently being intensely scrutinized due to its outstanding thermoelectric properties. Based on quantitative chemical bonding analysis we unravel that $Mg_3Sb_2$ exhibits a nearly isotropic three-dimensional (3D) bonding network with the interlayer and intralayer bonds being surprisingly similar, and these unique chemical bonding features are the origin of the nearly isotropic structural and thermal properties. The isotropic 3D bonding network is found to be broadly applicable to many Mg-containing compounds with the layered $CaAl_2Si_2$-type structure. Intriguingly, a parameter based on the electron density can be used as an indicator measuring the anisotropy of lattice thermal conductivity in layered structures. This work extends our understanding of structure and properties based on chemical bonding analysis, and it will guide the search for, and design of, layered materials with tailored anisotropic properties.

**Main Text:**

Chemical bonding, as the language of chemists, paves an intuitive shortcut for understanding the structure and properties of materials[1]. One notable example is two-dimensional (2D) layered transition metal dichalcogenides, where the main feature is the weak interlayer van der Waals interaction. Due to the weak van der Waals interaction, properties such as lattice thermal conductivity generally show strong anisotropy[2,3]. There are significant studies on quantification of weak van der Waals bonding in transition metal dichalcogenides[4,5], but the chemical bonding, especially the interlayer interaction, of many other layered materials remains largely unknown.

In recent years, $AB_2X_2$ compounds crystalizing in the layered $CaAl_2Si_2$ structure have attracted considerable research interest because of their promising magnetic and thermoelectric properties[6-12]. In particular, several compounds including n-type $Mg_3Sb_2$-based materials[13-16], and p-type $YbCd_{1.5}Zn_{0.5}Sb_2$[11], $EuZn_{1.75}Cd_{0.25}Sb_2$[12], and $Eu_{0.2}Yb_{0.2}Ca_{0.6}Mg_2Bi_2$[17] were discovered to show excellent thermoelectric figures of merit larger than unity. This type of structure covers an exceptionally rich variety of compounds, where A is an alkaline-earth or a divalent rare-earth element, B is a transition metal or a main group element, and X usually belongs to group 14 and 15[18]. In general, $AB_2X_2$ with the $CaAl_2Si_2$-type structure is understood as a Zintl phase by assuming that the covalent $B_2X_2$ layer receives the electrons donated by the ionic A layer[10,18-20]. This Zintl formulism has been very successful in explaining the electronic transport manipulation for promising thermoelectric materials[10,19]. It is commonly accepted that the interlayer A-X interaction is much weaker than the intralayer covalent bonding in the $B_2X_2$ layer[18,20]. The structural formation and its correlation with the electronic structure in $CaAl_2Si_2$ were studied in detail by different theoretical models based on $[Al_2Si_2]^{2-}$ networks with or without the effect of Ca cations[18,21-23]. In addition, electrical transport properties of $CaAl_2Si_2$- type compounds were rationalized by band structure engineering via an atomic orbital scheme[24]. Despite there are intensive theoretical studies on how crystal orbitals affect the electronic structure and electrical transport, there is very little knowledge on the quantitative description of chemical bonding, especially the interlayer interaction and how the chemical bonding affects thermal properties in $CaAl_2Si_2$-type compounds.

Here we report a quantitative analysis of chemical bonding in an archetypical compound $Mg_3Sb_2$ based on Bader's quantum theory of atoms in molecules[25], and compare it with the structurally related layered van der Waals solid $SnS_2$. It is found that $Mg_3Sb_2$ possesses a nearly isotropic 3D chemical bonding network with the interlayer bond being mostly ionic with partial covalent nature, and comparable to the intralayer interactions. Such a unique bonding feature in $Mg_3Sb_2$ not only challenges the well-known Zintl formalism, but also results in nearly isotropic thermal expansion coefficients, lattice compression, atomic displacement parameters, and lattice thermal conductivity. Importantly, we show how a simplified parameter based on the electron density can be used as an indicator for the anisotropy of the lattice thermal conductivity. Furthermore, the isotropic 3D chemical bonding network is found to be widely applicable to many other Mg-containing compounds with the $CaAl_2Si_2$-type structure.

$AB_2X_2$ with the $CaAl_2Si_2$-type structure can be described by tightly bound $[B_2X_2]^{2-}$ layers sandwiched by two-dimensional layers of $A^{2+}$ ions (Fig. 1a). Besides the interlayer A-X bond ($d_1$), two types of bonds exist in the $B_2X_2$ slabs, i.e. the tilted and vertical B-X bond, where the vertical bond ($d_3$) is often longer than the tilted bond ($d_2$)[18]. Three nonequivalent atoms have completely different coordination environments: A is connected to six X atoms with six equal bonds, B is tetrahedrally coordinated by X atoms with three tilted B-X bonds and one vertical B-X bond, and X is coordinated by three A atoms and four B atoms with seven adjacent bonds including three interlayer A-X bonds, three tilted B-X bonds, and one vertical B-X bond (Fig. 1c and Supplementary Fig. 1).

$Mg_3Sb_2$ (Space group: $P\bar{3}m1$, $a$ = 4.56187(3) and $c$ = 7.22944(6) Å at 299 K) can be considered as a special case of the $CaAl_2Si_2$ ($AB_2X_2$) structure in which A and B are Mg1 and Mg2, respectively. To have a better understanding of the interlayer interaction, a layered metal dichalcogenide $SnS_2$ with the trigonal $CdI_2$-type structure (Space group: $P\bar{3}m1$, $a$ = 3.6456(4) and $c$ = 5.8934(11) Å at 300 K)[26] was chosen for comparison since it shares many structural similarities with $Mg_3Sb_2$. Without considering the difference in lattice parameters, $Mg_3Sb_2$ can be viewed as intercalating two monolayers of Mg ions into the van der Waals gap of $SnS_2$ and replacing Sn and S respectively by Mg and Sb (Fig. 1b). Unlike the Sb atom surrounded by seven Mg atoms in $Mg_3Sb_2$, the S atom is coordinated by three Sn atoms and three S atoms with six adjacent bonds including three intralayer Sn-S bonds ($d_1'$) and three interlayer S-S bonds ($d_2'$) (see Fig. 1e).

The Non-Covalent Interaction (NCI) index, based on the electron density, $\rho$, and its derivatives, is a powerful tool to reveal weak interlayer interactions[27]. NCI analysis is based on the reduced density gradient (RDG) as a function of sign($\lambda_2$)$\rho$ (see methods), where sign($\lambda_2$) is the sign of the second eigenvalue of the electron density Hessian matrix[27,28]. Negative values of sign($\lambda_2$)$\rho$ indicate attractive interactions, whereas positive values suggest repulsive interaction. Spikes induced by the significant change in RDG approaching zero at critical points within low density regions correspond to weak interactions. The density value of the spike with low RDG relates to the strength of the corresponding interaction.

3D RDG isosurfaces with blue-green-red color (BGR) scales representing sign($\lambda_2$)$\rho$ values are given in Fig. 1d,f. Dark green RDG isosurfaces indicate that the interlayer Mg1-Sb in Mg$_3$Sb$_2$ is an attractive interaction, stronger than the weak interlayer S-S interaction in SnS$_2$ with RDG isosurfaces colored in green. In order to quantitatively understand the interlayer interactions, the dependence of RDG on sign($\lambda_2$)$\rho$ is calculated and shown in Fig. 1g,h. As expected, distinct differences can be seen between the weak interlayer S-S and intralayer Sn-S interactions in SnS$_2$. Compared with the intralayer Sn-S interaction, interlayer S-S interaction shows a low RDG peak with a much smaller sign($\lambda_2$)$\rho$ value approaching zero, a clear indication of weak van der Waals interaction (Fig. 1h). In contrast, the RDG distribution of the interlayer Mg1-Sb interaction in Mg$_3$Sb$_2$ is very similar to those of the intralayer Mg2-Sb interactions (Fig. 1g). The density value of the low RDG peak for the interlayer Mg1-Sb interaction is just slightly lower than those of the vertical and tilted Mg2-Sb interlayer interactions. This indicates that the interlayer and intralayer interactions in Mg$_3$Sb$_2$ are comparable; that is, the tilted Mg2-Sb bond is slightly stronger than the vertical Mg2-Sb bond, while the vertical Mg2-Sb bond is slightly stronger than the interlayer Mg1-Sb interaction.

Static deformation electron density maps of the (110) planes in Mg$_3$Sb$_2$ and SnS$_2$ are shown in Fig. 1i,j. The (110) plane is chosen because all nonequivalent bonds are included in this plane. Interestingly, in addition to the expected charge accumulations along intralayer Mg2-Sb bonds, clear charge accumulation is observed along the interlayer Mg1-Sb bond in Mg$_3$Sb$_2$. The Sb atom possesses seven valence shell charge concentrations (VSCC) towards Mg atoms, including three VSCC towards Mg1 atoms and four towards Mg2 atoms (Figs. 1i and 1c). Comparing with the static deformation density map of SnS$_2$, the charge accumulation along interlayer Mg1-Sb is not surprising since interlayer Mg1-Sb ($d_1$) bond in Mg$_3$Sb$_2$ corresponds to the intralayer Sn-S bond ($d_1'$) in SnS$_2$. In spite of similar features of density deformation profiles between Mg1-Sb and Sn-S, much larger charge accumulation and density deformation can be seen along the Sn-S bond in SnS$_2$ (Fig. 1j). This implies that the interlayer Mg1-Sb can be viewed as a weakened form of the intralayer Sn-S bond with longer bond length and less covalent nature.

Topological analysis of the theoretical full electron density (ED) was conducted based on Bader's quantum theory of atoms in molecules[25] (Supplementary Note 1 and Supplementary Figs. 1 and 2). Topological properties at the bond critical points (BCPs) are provided in Table 1. In Mg$_3$Sb$_2$, BCPs are found close to the Mg atoms along the vertical and tilted Mg2-Sb bonds and along Mg1-Sb bond (see Supplementary Fig. 1). If only based on the sign of the Laplacian $\nabla^2\rho(\mathbf{r}_b)$ at BCPs, the interlayer Mg1-Sb bond together with the two intralayer Mg2-Sb bonds can be described as closed-shell interactions or ionic bonds[29] (see also Supplementary Fig. 3). However, the nature of chemical bonds cannot simply be judged by the sign of $\nabla^2\rho(\mathbf{r}_b)$. A more accurate analysis is using the Laplacian profile in comparison with that of the Independent Atom Model (IAM). As can be seen in Supplementary Fig. 4, for all three types of bonds in Mg$_3$Sb$_2$ the Laplacian values along the bond path are less positive than those of IAM. This indicates the partial covalent nature (high polarity) in these bonds. Considering the degree of differences in Laplacian profiles, the proportion of partial covalency is slightly increasing from the interlayer Mg1-Sb to the vertical Mg2-Sb to the tilted Mg2-Sb.

According to the well-established classification scheme[29], the interlayer Mg1-Sb and intralayer Mg2-Sb bonds can be described as polar bonds according to the BCPs properties with the small density $\rho(\mathbf{r}_b)$, positive $\nabla^2\rho(\mathbf{r}_b)$, $|V|/G$ being slightly larger than unity, negative total energy density $H$, and $G/\rho < 1$ (see Table 1). Changing from the tilted Mg2-Sb to the vertical Mg2-Sb to the Mg1-Sb bond, the density value undergoes a negligible decrease, which indicates a minor decrease in covalency and interaction strength. In contrast, a remarkable difference is observed between the interlayer S-S and intralayer Sn-S bonds in SnS$_2$ (see Table 1). The interlayer S-S interaction can be described as a weak van der Waals bond as judged by the very small electron density $\rho(\mathbf{r}_b)$, positive $\nabla^2\rho(\mathbf{r}_b)$, $|V|/G < 1$, and positive $H$ at the BCP, whereas the intralayer Sn-S interaction can be treated as a polar covalent bond based on the positive $\nabla^2\rho(\mathbf{r}_b)$, $1 < |V|/G < 2$, negative $H$, and $G/\rho < 1$. The electron density at the BCP of the interlayer S-S bond is approximately 8 times smaller than that of the Sn-S bond, indicating the much weaker strength of the interlayer interaction compared with that of the intralayer interaction. Upon comparison of topological properties at the BCPs (see Table 1), the three similar bonds in Mg$_3$Sb$_2$ are clearly more polar and weaker than the Sn-S bond in SnS$_2$, but they are stronger than the weak interlayer S-S interaction. The above results are consistent with the NCI analysis.

The topological properties at BCPs of another archetypical CaAl$_2$Si$_2$-type compound, CaZn$_2$Sb$_2$, are analyzed and compared with those of Mg$_3$Sb$_2$ (see Table 1). Clearly, the interlayer interactions are quantitatively similar, but the intralayer interactions in B$_2$X$_2$ slabs show significant differences. The intralayer Zn-Sb bonds in CaZn$_2$Sb$_2$, described as polar covalent bonds, show much larger values of $\rho(\mathbf{r}_b)$, $|V|/G$, and $-H$, which denotes that the Zn-Sb bonds are much more covalent and stronger than the interlayer Ca-Sb bond and the Mg2-Sb bonds in Mg$_3$Sb$_2$. This is consistent with the smaller electronegativity difference and shorter bond distance between Zn and Sb. Despite the stronger Zn-Sb interaction, the difference between the interlayer and intralayer interactions in CaZn$_2$Sb$_2$ is moderate if compared to that of SnS$_2$.

The above deduction is further strengthened by the values of the Bader atomic properties shown in Table 2. The atomic charges of Zn and Sb in CaZn$_2$Sb$_2$ being far from the nominal oxidation states suggest a high degree of covalency in the [Zn$_2$Sb$_2$]$^{2-}$ slabs, whereas the large charge transfer of the Ca atom indicates largely ionic features in the $Ca^{2+}$ layers. In contrast, nearly complete charge transfers are observed for all atoms in $Mg_3Sb_2$ with Mg1 and Mg2 showing nearly the same atomic charges, which indicates that the interlayer and intralayer bonds are largely ionic and comparable. Based on the above results, the argument that the ionic $A^{2+}$ layer donates electrons to the covalent $[B_2X_2]^{2-}$ layer holds true for $CaZn_2Sb_2$; however, this argument is not applicable to $Mg_3Sb_2$ since the intralayer bonds in $[Mg_2Sb_2]$ slabs are not really covalent. Therefore, since the chemical bonds in $Mg_3Sb_2$ are largely ionic without a clear distinction of ionic and covalent parts we approach the limit of the application of the well-recognized Zintl formalism in $CaAl_2Si_2$-type compounds.

The comprehensive chemical bonding analysis paves the way to understand the structure and properties. The largely ionic feature with partial covalency (high polarity) of chemical bonds in $Mg_3Sb_2$ explains the intrinsically poor carrier density and mobility as well as the reasonably low lattice thermal conductivity[30]. Furthermore, the comparable interlayer and intralayer interactions unveil the three-dimensional chemical bonding network in $Mg_3Sb_2$, ruling out a description as a typical 2D layered structure. Importantly, the 3D bonding network is nearly isotropic in $Mg_3Sb_2$ upon quantitative comparisons of topological properties of different bonds. Such a feature is decisive for many unique properties including structural parameters and thermal properties.

Chemical bonding analysis is crucial for understanding the lattice response under physical conditions such as temperature and pressure. Figure 2a shows the temperature dependence of the experimental lattice parameters of $Mg_3Sb_2$ (see also Supplementary Fig. 5 and Supplementary Table 1). As illustrated in the figure, the lattice parameters all display linear increasing trends as the temperature increases and the thermal expansion along the $c$ axis is slightly larger than that along the $a$ axis. The linear thermal expansion coefficients at room temperature along the $a$ and $c$ directions in $Mg_3Sb_2$ are respectively $1.88\times10^{-5}$ and $2.42\times10^{-5}$ K$^{-1}$, which leads to a nearly isotropic $\alpha_c/\alpha_a$ of 1.29, much smaller than those of typical layered materials $TiS_2$[31], $MoS_2$[32], $MoSe_2$[32], and $Bi_2Te_3$[33] (see Fig. 2b and Supplementary Table 2). Furthermore, similar results are observed in the pressure-induced lattice compression. The relative lattice parameters and interlayer distance as a function of a series of hydrostatic pressures are simulated using DFT calculations and given in Fig. 2c,d. Under the same pressure, the decrease of the relative lattice parameter $c/c_0$ in $Mg_3Sb_2$ is just slightly larger than that of $a/a_0$, whereas the lattice parameter $c$ is much more compressible than $a$ in all layered van der Waals solids $SnS_2$[26], $TiS_2$, $MoS_2$, and $MoSe_2$.

It is clear that both the lattice expansion with temperature and the lattice compression under pressure exhibit nearly isotropic features in $Mg_3Sb_2$, which can be essentially understood by the 3D chemical bonding network in this material. In both cases, the lattice parameter $c$ in $Mg_3Sb_2$ shows slightly larger thermal expansion or pressure compression than that of $a$, which can be attributed to the slightly weaker interlayer bond compared with the intralayer bonds. Moreover, the interlayer distance of $Mg_3Sb_2$ is less compressible than those of van der Waals solids, confirming the interlayer interaction being stronger than the van der Waals force.

To gain insight on how chemical bonding affects the thermal motion of the atoms, the isotropic atomic displacement parameters $U_{iso}$ of $Mg_3Sb_2$ were obtained from Rietveld refinement of multi-temperature synchrotron powder X-ray diffraction (PXRD) data (Fig. 3a). Refinement details are provided in Supplementary Information (see Supplementary Table 1 and Supplementary Note 2). As illustrated in Fig. 3b, the Mg1 atoms exhibit larger thermal displacements than those of the Mg2 and Sb atoms. This experimental trend is well reproduced by the theoretical result based on the harmonic approximation shown in Fig. 3c. The atomic thermal motions are closely related to the potential energy surfaces (see the inset of Fig.3c). By displacing the atoms from their equilibrium positions, we found that the Mg1 atom shows a relatively flat potential well and it is thereby loosely bonded, consistent with its relatively large thermal vibration. In fact, the larger thermal displacement and flatter potential originate from the weaker adjacent bonds (i.e., the interlayer Mg1-Sb bonds) of the Mg1 atom. In addition, a slightly larger $U_{iso}$ of Mg2 than that of Sb in the $[Mg_2Sb_2]^{2-}$ slabs was observed at elevated temperatures, which can be rationalized by the difference in atomic masses.

When considering the atomic displacement parameters and potential wells along the axial directions, both the Sn and S atoms in layered $SnS_2$ display highly anisotropic characteristics with the vibration along the $c$ direction being significantly larger than along the $a$ direction (see Fig. 3d and the inset of Fig. 3d). This is due to the weak van der Waals interaction along the $c$ axis. However, the atomic displacement parameters and potential wells of all atoms in $Mg_3Sb_2$ are relatively isotropic along different axial directions (Fig. 3c). The Mg1 atom manifests perfect isotropic features due to its equally adjacent Mg1-Sb bonds, while the Mg2 and Sb atoms show less isotropic features in atomic displacement parameters because of their comparable but nonequivalent adjacent bonds. The above result is another validation for the notion of comparable interlayer and intralayer bonding interactions in $Mg_3Sb_2$. Furthermore, it is found that the potential wells of all atoms except Mg2 and Sb along the $c$ direction are ideally harmonic, which can be reasonably understood by the atoms being surrounded by symmetric electron density along the axial directions (see Supplementary Figs. 6a and 7). However, slightly anharmonic features can be seen in potential wells of the Mg2 and Sb atoms along the *c* direction (see Supplementary Fig. 6b), which is induced by the different strengths of the three nonequivalent bonds in $Mg_3Sb_2$.

It is well known that weak chemical bonding usually accompanied by strong lattice anharmonicity leads to low lattice thermal conductivity[34,35]. To probe the effect of chemical bonding on thermal transport, the lattice thermal conductivity was simulated by DFT calculations. Indeed, weak van der Waals interaction between the layers in $SnS_2$ leads to considerably lower lattice thermal conductivity along the *c* axis in comparison with that along the *a* axis (see Fig. 4a). High anisotropic ratio $\kappa_a/\kappa_c$ of lattice thermal conductivity is commonly observed at 300 K in typical layered materials, such as 15.8 in $SnS_2$, 16.2 in $TiS_2$, 16.2 in $MoS_2$[3], and 11.2 in $MoSe_2$[3] (Fig. 4b). In contrast, unlike the noticeable anisotropy in van der Waals solids, the lattice thermal conductivity in $Mg_3Sb_2$ is nearly isotropic with $\kappa$ along the *c* axis being negligibly lower than that along the *a* axis ($\kappa_a/\kappa_c \approx 1.1$ at 300 K, see Fig. 4).

In order to elucidate the origin of the isotropic lattice thermal conductivity in $Mg_3Sb_2$, phonon dispersion, group velocity, and Grüneisen parameter were calculated. The Grüneisen parameter, defined as the response of phonon frequencies to volume change, represents the strength of the lattice anharmonicity. The average Grüneisen parameters along the *a* and *c* axes in $Mg_3Sb_2$ are 1.8 and 2.2, respectively, which gives an anisotropic ratio of ~1.2, much smaller than that of ~3.0 in layered $SnS_2$ (see the inset table of Fig. 4a). The slightly higher Grüneisen parameter, as well as the aforementioned weak anharmonic potential wells along the *c* axis in $Mg_3Sb_2$ induced by the slightly weaker interlayer Mg1-Sb bond, explains the smaller lattice thermal conductivity along this direction. In addition to the Grüneisen parameter, the phonon dispersion and group velocity also show nearly isotropic features in $Mg_3Sb_2$, whereas those in layered SnS$_2$ are considerably anisotropic (see Supplementary Fig. 8 and Supplementary Table 3).

The origin of anisotropy in thermal properties can be traced to the chemical bonding. We can define a simplified parameter, the intralayer-to-interlayer bond-strength ratio $\rho_{intra}/\rho_{inter}$, which measures the degree of anisotropy of the chemical bonding network in a layered structure. $\rho_{intra}$ and $\rho_{inter}$ denote the electron density values at BCPs of the intralayer and interlayer bonds, respectively. For the layered AB$_2$X$_2$ compounds with two nonequivalent intralayer bonds, $\rho_{intra}$ is calculated by averaging the electron density values at BCPs of the two intralayer bonds. As can be seen in Fig. 4b, a nearly linear correlation between the anisotropic ratio $\kappa_a/\kappa_c$ of lattice thermal conductivity and $\rho_{intra}/\rho_{inter}$ is revealed. This suggests that $\rho_{intra}/\rho_{inter}$ can be adopted as an indicator for the anisotropy of lattice thermal conductivity. $\rho_{intra}/\rho_{inter} \approx 1$ in Mg$_3$Sb$_2$ indicates a nearly isotropic 3D chemical bonding network, which results in the nearly isotropic feature in phonon dispersion, group velocity, Grüneisen parameter, and eventually in lattice thermal conductivity.

It should be noted that the nearly isotropic 3D bonding network is not limited to Mg$_3$Sb$_2$. The topological properties of several other Mg-containing compounds including Mg$_3$Bi$_2$, CaMg$_2$Sb$_2$, CaMg$_2$Bi$_2$, SrMg$_2$Sb$_2$, and YbMg$_2$Sb$_2$ are shown in Supplementary Tables 4-6. All these compounds show comparable interlayer and intralayer polar bonds with $\rho_{intra}/\rho_{inter} \approx 1$, similar to those of Mg$_3$Sb$_2$. This suggests that the nearly isotropic 3D bonding network is a general feature in Mg-containing compounds with the layered CaAl$_2$Si$_2$ structure.

In summary, using quantitative analysis of chemical bonding, we have shown that the interlayer interaction in Mg$_3$Sb$_2$ is largely ionic with partial covalent nature, and it exhibits the same type of interaction with comparable strength as the intralayer chemical bonds. This result challenges the widely accepted Zintl concept that assumes the $[Mg_2Sb_2]^{2-}$ slabs being covalently bonded. The nearly isotropic 3D bonding network formed by the comparable chemical bonds leads to the isotropic characteristics in many properties, such as lattice thermal expansion, lattice compression under hydrostatic pressure, atomic displacement parameters, and lattice thermal conductivity. Interestingly, the intralayer-to-interlayer bond-strength ratio based on the electron density is established as a simplified descriptor for the anisotropy of lattice thermal conductivity. Moreover, it is found that the nearly isotropic 3D chemical bonding network is not limited to $Mg_3Sb_2$ but can be broadly applied to many other Mg-containing materials with the $CaAl_2Si_2$ structure. Thus, this work extends our fundamental understanding of the structure-property relationship using chemical bonding as a bridge and it will guide the rational design of layered materials with tailored properties.

**Methods**

**Sample synthesis.** $Mg_3Sb_2$ crystals were synthesized from the mixture of high-purity elements Sb pieces (99.9999%, Chempur) and Mg turnings (99.5%, Chempur). 0.8 mole excess Mg was added to compensate the evaporation loss during the synthesis. The mixture of Mg and Sb was loaded into a glassy carbon crucible with a lid. The crucible was placed in a quartz tube, which was evacuated to a pressure smaller than $10^{-4}$ mbar and flame-sealed. $Mg_3Sb_2$ crystals were obtained using vertical Bridgman crystal growth by heating to 973 K for 24 h and then slowly cooling down to room temperature over 160 h with the sample moving at a rate of 2 mm h$^{-1}$.

**Synchrotron powder X-ray diffraction.** $Mg_3Sb_2$ powders, obtained by crushing the crystals, were floated in ethanol in order to select very small and homogenous particles. The obtained fine powders were packed under Ar in a 0.2 mm diameter quartz capillary. Synchrotron PXRD patterns were collected at SPring-8 BL44B2 beamline[36] with a wavelength of 0.500197(1) Å, which was calibrated by a CeO$_2$ standard. Data were collected at 299-813 K (heating) and 770-299 K (cooling) with the high temperature setup (see Supplementary Fig. 9). Synchrotron PXRD patterns were analyzed using Rietveld refinement in the FullProf program[37]. The peak profiles were described by the Thompson-Cox-Hastings pseudo-Voigt function[38] convoluted with axial divergence asymmetry. The background was modeled using a linear interpolation between a set of manually selected background points. The detailed results of the refinements of all heating and cooling data were provided in Supplementary Tables 1, 2, and 7-9. The analysis of the data at 770-299 K upon cooling was used for discussion in the main text.

**Theoretical calculations.** Density functional theory (DFT) calculations in this work were performed in the Wien2k code[39] using a full-potential linear augmented plane-wave plus local orbitals method and with the VASP code[40] based on the projector-augmented wave method[41]. The structural parameters including lattice parameters and atomic positions were fully relaxed until the Hellmann-Feynman force criterion of 0.001 eV Å$^{-1}$ was reached in the VASP code. All calculations for Mg$_3$Sb$_2$, CaZn$_2$Sb$_2$, and other compounds with the CaAl$_2$Si$_2$-type structure were performed using the PBE functional[42], whereas the calculations for layered SnS$_2$, TiS$_2$, MoS$_2$, and MoSe$_2$ were carried out using van der Waals functional optB86b-vdW[43]. Here we use optB86b-vdW functional for metal dichalcogenides since it gives structural parameters under hydrostatic pressure in good agreement with those from experiment[26]. Structural parameters of Mg$_3$Sb$_2$, TiS$_2$, MoS$_2$, and MoSe$_2$ under hydrostatic pressure were calculated in this work, while the high-pressure data of SnS$_2$ is from reported experimental work[26]. An energy convergence criterion of $10^{-6}$ eV and a plane wave energy cutoff of 500 eV were adopted for calculations.

The electron density calculations were done with the Wien2k code on a dense 22×22×12 k mesh with the plane wave cutoff parameter $R_{MT}K_{max}$ of 10. The relaxed structural parameters were used for the calculations. The charge density inside the atomic spheres was expanded to form spherical harmonics up to $l_{max} = 10$. The muffin-tin radii ($R_{MT}$) for Mg, Sb, Sn, and S were chosen as 2.3, 2.5, 2.0, and 1.8 bohr, respectively. The topological analysis of the total electron density based on Bader's quantum theory of atoms in molecules was conducted with the Critic2 code[44,45]. Atomic charges and atomic basin volumes were calculated using the QTREE algorithm[46] as implemented in the Critic2 code. Non-covalent interaction (NCI) plots were computed with the NCIPLOT program[27,28]. NCI analysis is based on the electron density $\rho$ and its reduced density gradient (RDG). The RDG can be expressed as[27,28]

$$\text{RDG} = \frac{1}{2(3\pi^2)^{1/3}} \frac{|\nabla \rho|}{\rho^{4/3}}. \tag{1}$$

The 3D NCI plot with RDG isosurfaces in real space was visualized by VMD[47]. The isosurfaces were colored according to the value of sign($\lambda_2$)$\rho$, where a BGR color scale was adopted. Blue color represents attractive or bonding interaction, green weak van der Waals interaction, and red repulsive interaction. To have a clear view of the interlayer and intralayer interactions, the isosurfaces with sign($\lambda_2$)$\rho > 0$ are eliminated in Fig. 1d,f.

The harmonic phonon dispersion and atomic displacement parameters were computed with VASP and Phonopy[48] using the finite displacement method[49]. The default displacement amplitude of 0.01 Å was used for $Mg_3Sb_2$ and $SnS_2$. The calculations were done in 4×4×2 supercells to balance the computational cost and well-converged phonon dispersions (see Supplementary Fig. 10). The results of phonon dispersions for $Mg_3Sb_2$ and $SnS_2$ are shown in Supplementary Fig. 8. ±5% of the equilibrium volume was used for the calculation of mode

Grüneisen parameter. The average Grüneisen parameter was calculated using

$$\overline{\gamma} = \frac{\sum\limits_{\mathbf{q},i} \gamma(\mathbf{q},i) C_V(\mathbf{q},i)}{\sum\limits_{\mathbf{q},i} C_V(\mathbf{q},i)}, \tag{2}$$

where $\gamma(\mathbf{q},i)$ is the mode Grüneisen parameter for the phonon branch $i$ at wave vector $\mathbf{q}$, given as

$$\gamma(\mathbf{q},i) = -\frac{V}{\omega(\mathbf{q},i)} \frac{\partial \omega(\mathbf{q},i)}{\partial V}, \tag{3}$$

where $\omega(\mathbf{q},i)$ is phonon frequency, $V$ is volume, and $C_V(\mathbf{q},i)$ is the mode dependent heat capacity defined as

$$C_V(\mathbf{q},i) = k_B \left( \frac{\hbar \omega(\mathbf{q},i)}{k_B T} \right)^2 \frac{e^{\hbar \omega(\mathbf{q},i)/k_B T}}{(e^{\hbar \omega(\mathbf{q},i)/k_B T} - 1)^2}. \tag{4}$$

Another averaging method of Grüneisen parameter and the corresponding result are shown in Supplementary Note 3 and Supplementary Table 10. Moreover, group velocities along the axial directions for the acoustic branches were computed and they are shown in Supplementary Table 3. The lattice thermal conductivity was computed with the ShengBTE code based on a full iterative solution to the Boltzmann transport equation for phonons[50]. The second-order interatomic force constants were computed in the 4×4×2 supercells. Considering the computational cost, the third-order interatomic force constants were calculated in the 4×4×2 supercells for $SnS_2$ and the 3×3×3 supercells for $Mg_3Sb_2$, $CaMg_2Sb_2$, $CaMg_2Bi_2$, $CaZn_2Sb_2$, and $SrZn_2Sb_2$. To ensure well-converged thermal conductivity (see Supplementary Fig. 11), the interaction range up to the seventh nearest neighbors was considered for the calculations of the third-order interatomic force constants. The calculation details of lattice thermal conductivity of another metal dichalcogenide $TiS_2$ are also provided in Supplementary Note 2 and Supplementary Fig. 12.

**Acknowledgments**

This work was supported by the Danish National research Foundation (Center for Materials Crystallography, DNRF93), the Danish Center for Synchrotron and Neutron Research (Danscatt), and the Danish Center for Scientific Computing. The authors would like to thank the synchrotron beamline RIKEN BL44B2 (Proposal No. 20160037) at SPring-8 for the beamtime allocation. The numerical results presented in this work were obtained at the Center for Scientific Computing, Aarhus.

**Author contributions**

J.Z. and B.B.I. designed the study. L.S. and J.Z. synthesized the samples and characterized structures. J.Z. performed theoretical calculations. M.S. and K.T. provided discussions. J.Z. and B.B.I. wrote the manuscript. All other authors read and edited the manuscript.

**Author Information**

Correspondence and requests for materials should be addressed to B.B.I. (bo@chem.au.dk).

**Competing financial interests**

The authors declare no competing financial interests.

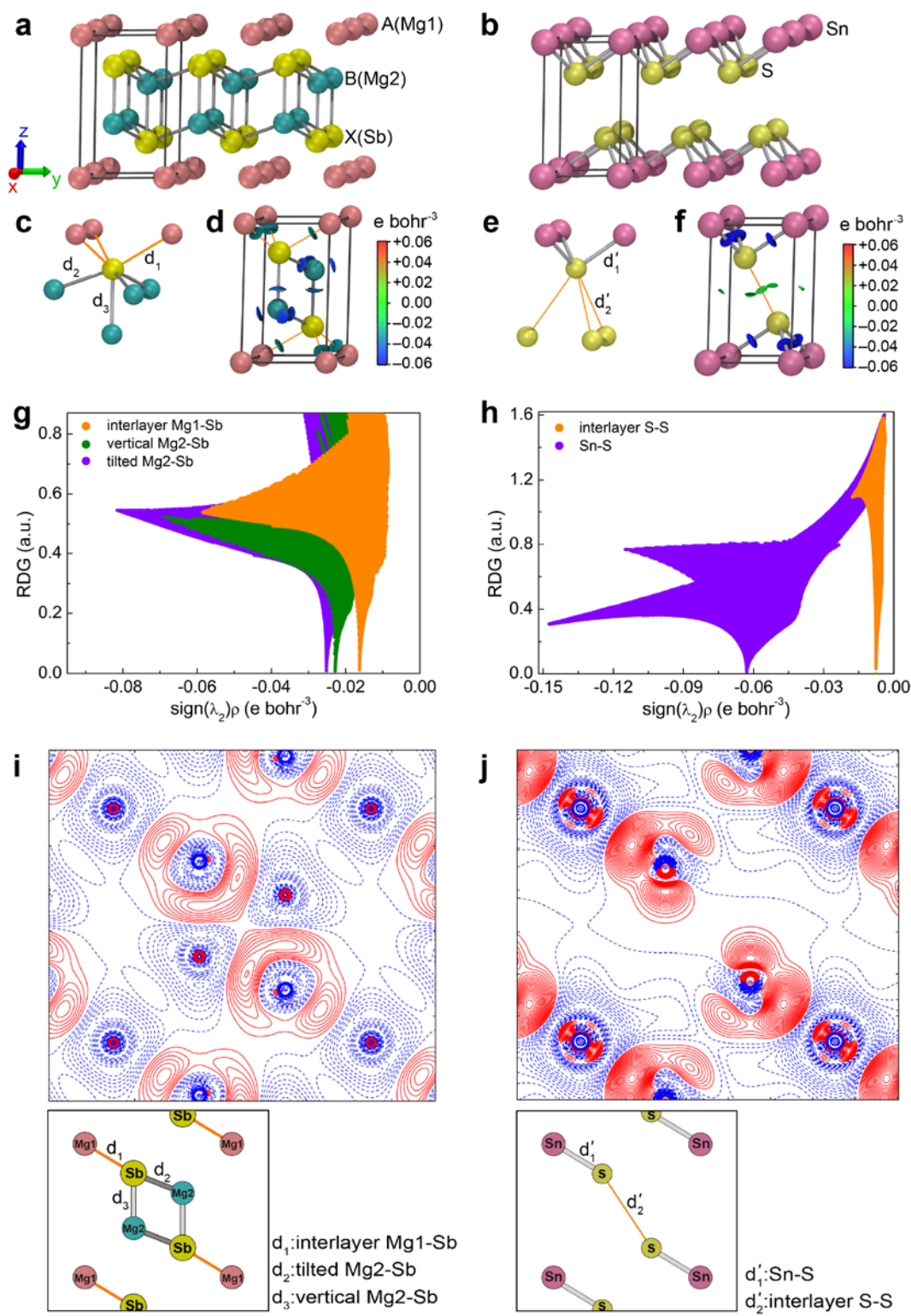

**Figure 1 | Crystal structure, non-covalent interaction analysis, and static deformation electron density.** (**a, b**) Crystal structure of (a) $AB_2X_2$ ($Mg_3Sb_2$) and (b) $SnS_2$. Mg1 and Mg2 represent Mg atoms in the Mg monolayer and $[Mg_2Sb_2]^{2-}$ layer, respectively. (**c, e**) Coordination polyhedron of (c) the X (Sb) atom in $AB_2X_2$ ($Mg_3Sb_2$) and (e) the S atom in $SnS_2$. (**d, f**) Non-

covalent interaction plot for (d) Mg$_3$Sb$_2$ and (f) SnS$_2$. The RDG isosurface corresponds to RDG = 0.24 a.u., which is colored on a BGR scale of -0.06 < sign($\lambda_2$)$\rho$ < 0.06 $e$ bohr$^{-3}$. (**g, h**) RDG as a function of sign($\lambda_2$)$\rho$ for the interlayer and intralayer interactions in (g) Mg$_3$Sb$_2$ and (h) SnS$_2$. (**i, j**) Static deformation map on (110) plane containing both interlayer and intralayer interactions of (i) Mg$_3$Sb$_2$ and (j) SnS$_2$. The contour interval is 0.006 $e$ Å$^{-3}$. Positive (negative) contours are plotted with red full (blue dotted) lines. The inset shows the corresponding (110) plane.

**Table 1 | Topological properties of the bond critical points ($r_b$).** $d$ is the bond length. $\rho(r_b)$ and $\nabla^2\rho(r_b)$ are the charge density and its Laplacian at the BCP, respectively. $G$ and $V$ denote the kinetic and potential energy at the BCP, respectively. $H$ is the total energy density ($H = G + V$). $G$, $V$, $H$ and $G/\rho$ are in a.u.

| Bond | $d$ (Å) | $\rho(r_b)$ ($e$ Å$^{-3}$) | $\nabla^2\rho(r_b)$ ($e$ Å$^{-5}$) | $G$ (a.u.) | $V$ (a.u.) | $H$ (a.u.) | $|V|/G$ | $G/\rho$ (a.u.) |
|---|---|---|---|---|---|---|---|---|
| Mg$_3$Sb$_2$ | | | | | | | | |
| Interlayer Mg1-Sb | 3.120 | 0.110 | 0.649 | 0.0075 | –0.0082 | –0.0008 | 1.101 | 0.460 |
| Tilted Mg2-Sb | 2.849 | 0.170 | 1.267 | 0.0150 | –0.0168 | –0.0018 | 1.122 | 0.595 |
| Vertical Mg2-Sb | 2.959 | 0.154 | 0.997 | 0.0122 | –0.0140 | –0.0018 | 1.150 | 0.533 |
| CaZn$_2$Sb$_2$ | | | | | | | | |
| Interlayer Ca-Sb | 3.230 | 0.122 | 0.831 | 0.0093 | –0.0100 | –0.0007 | 1.074 | 0.516 |
| Tilted Zn-Sb | 2.719 | 0.313 | 0.734 | 0.0223 | –0.0370 | –0.0147 | 1.659 | 0.480 |
| Vertical Zn-Sb | 2.820 | 0.259 | 0.738 | 0.0177 | –0.0277 | –0.0100 | 1.567 | 0.460 |
| SnS$_2$ | | | | | | | | |
| Interlayer S-S | 3.609 | 0.051 | 0.485 | 0.0042 | –0.0034 | 0.0008 | 0.801 | 0.553 |
| Sn-S | 2.592 | 0.424 | 1.512 | 0.0390 | –0.0624 | –0.0233 | 1.598 | 0.620 |

**Table 2 | Atomic properties.** $Q$ and $V$ represent the atomic charge and the atomic basin volume, respectively.

| Atoms | $Q$ | $V$ (Å$^3$) |
|---|---|---|
| Mg$_3$Sb$_2$ | | |
| Mg1 | 1.51 | 8.71 |
| Mg2 | 1.47 | 8.50 |
| Sb | –2.23 | 53.69 |
| CaZn$_2$Sb$_2$ | | |
| Ca | 1.37 | 15.71 |
| Zn | 0.31 | 18.74 |
| Sb | –0.99 | 38.85 |
| SnS$_2$ | | |
| Sn | 1.53 | 15.40 |
| S | –0.76 | 26.81 |

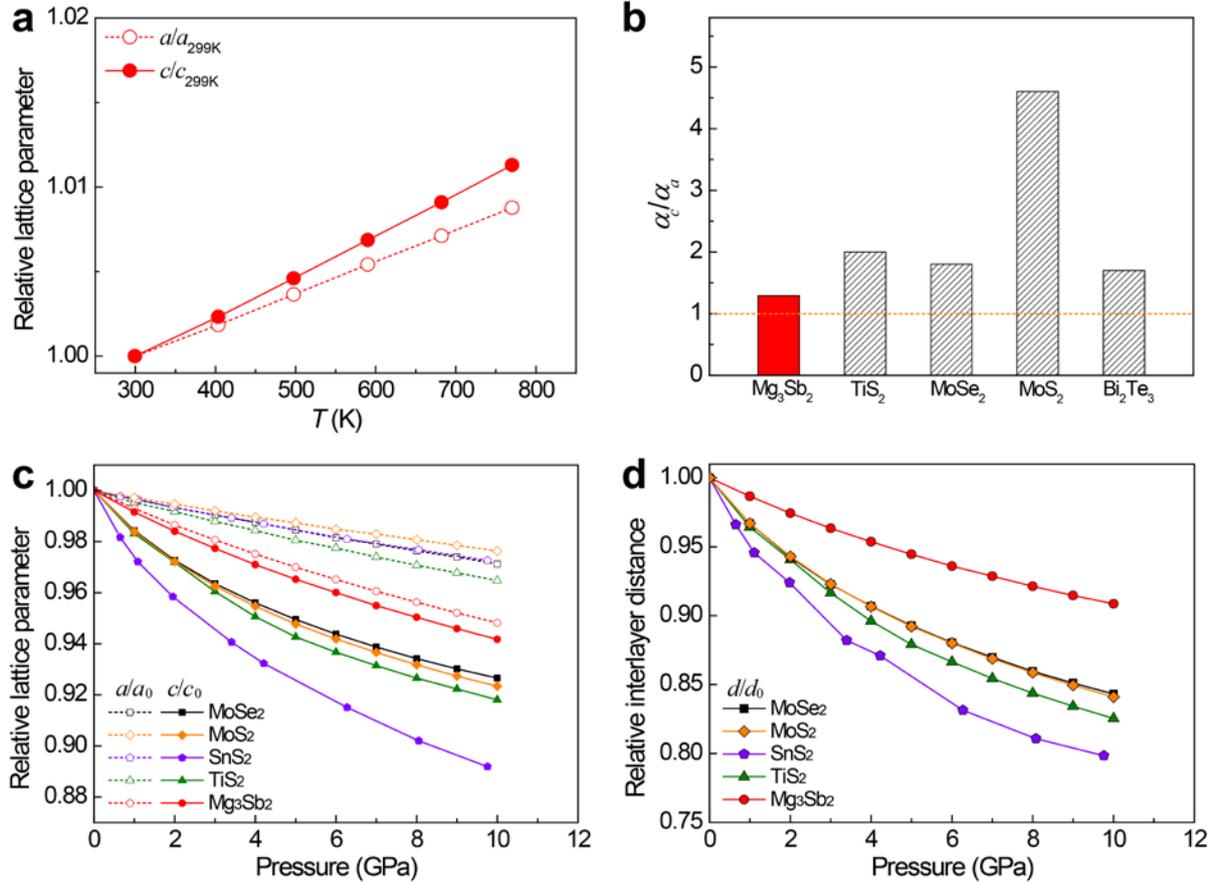

**Figure 2 | Relative lattice parameters under thermal expansion and high-pressure compression.** (a) Relative lattice constants $a/a_{299K}$ and $c/c_{299K}$ of $Mg_3Sb_2$ as a function of temperature. The data is obtained from Rietveld refinement of the multi-temperature synchrotron PXRD. (b) Anisotropic ratio of the linear thermal expansion coefficient $\alpha_c/\alpha_a$ in $Mg_3Sb_2$ compared to those of several typical layered materials, that is, $TiS_2$[31], $MoSe_2$[32], $MoS_2$[32], and $Bi_2Te_3$[33]. (c, d) The calculated pressure dependence of (c) the relative lattice parameters and (d) the relative interlayer distance of $Mg_3Sb_2$ in comparison with those of several typical layered metal dichalcogenides. The data of $SnS_2$ are adapted from the reported experimental work[26].

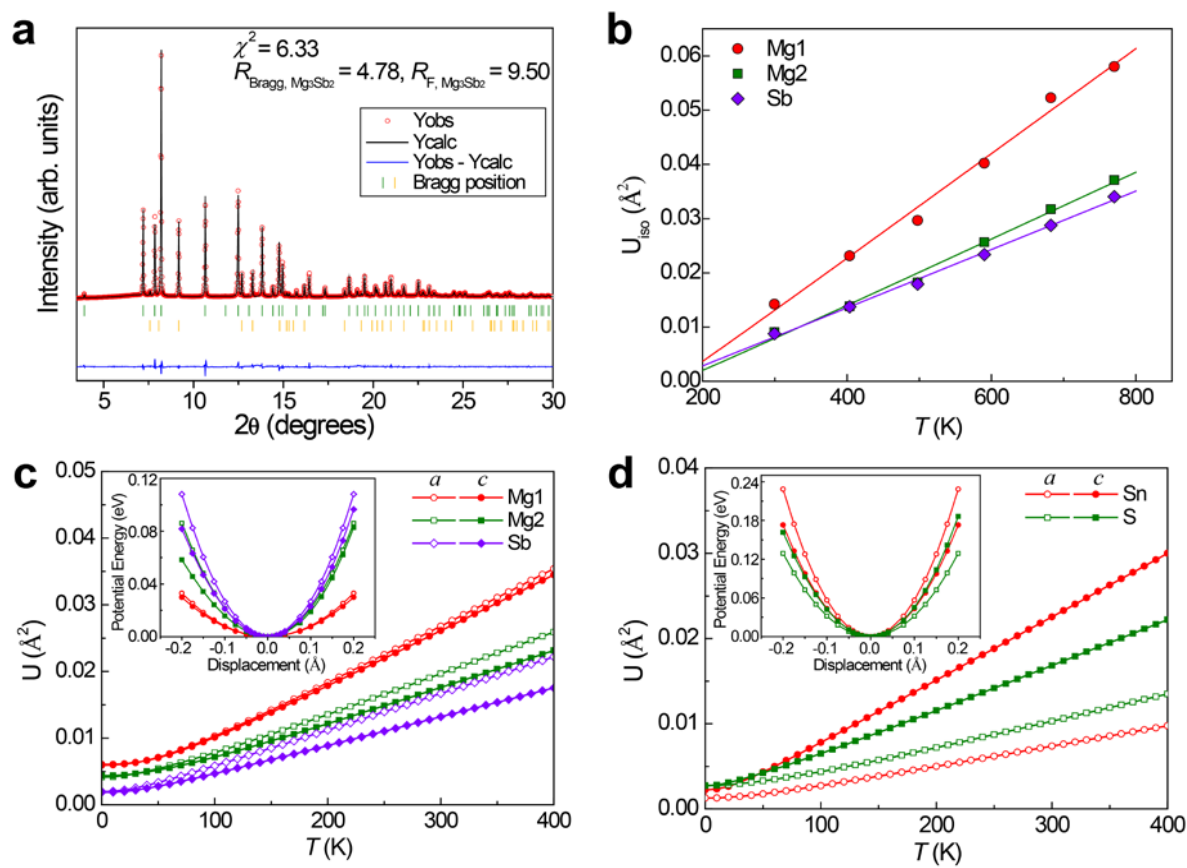

**Figure 3 | Atomic displacement parameters. (a)** Calculated (Rietveld method) and observed synchrotron PXRD patterns of $Mg_3Sb_2$ at 770 K upon cooling. Red open circles and black line are observed and calculated data, respectively. The blue line represents the difference between the observed and calculated patterns. The green and orange vertical bars correspond to the Bragg positions of the main phase $Mg_3Sb_2$ and the secondary phase Sb, respectively. **(b)** Temperature dependence of the experimental isotropic atomic displacement parameters of $Mg_3Sb_2$. **(c, d)** Temperature dependence of the theoretical atomic displacement parameters along different axial directions of (c) $Mg_3Sb_2$ and (d) $SnS_2$. The inset shows the potential energy curves for the nonequivalent atoms as a function of the displacements from the equilibrium positions.

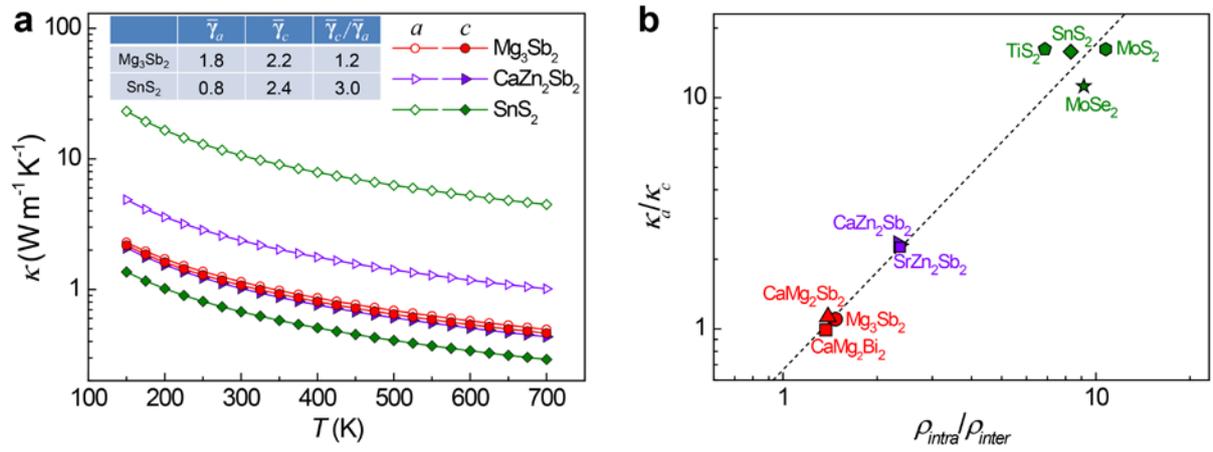

**Figure 4 | Lattice thermal conductivity. (a)** Calculated lattice thermal conductivity of $Mg_3Sb_2$, $CaZn_2Sb_2$, and $SnS_2$ along *a* and *c* directions as a function of temperature. The inset table displays the average Grüneisen parameters at room temperature along *a* and *c* directions. **(b)** Anisotropy of lattice thermal conductivity $\kappa_a/\kappa_c$ at 300 K as a function of the intralayer-to-interlayer bond-strength ratio $\rho_{intra}/\rho_{inter}$. $\rho_{intra}$ and $\rho_{inter}$ represent the electron density values at BCPs of the intralayer and interlayer bonds, respectively. The theoretical lattice thermal conductivity data of $MoSe_2$ and $MoS_2$ are adapted from ref. 3. The dashed line is a guide to the eyes.

# Supplementary Information for

# Relating chemical bonding to physical properties: The origin of unexpected isotropic properties in layered materials

*Jiawei Zhang, Lirong Song, Mattia Sist, Kasper Tolborg, and Bo Brummerstedt Iversen[*]*

*Center for Materials Crystallography, Department of Chemistry and iNANO, Aarhus University, DK-8000 Aarhus, Denmark*

*Corresponding author: bo@chem.au.dk

**Including:**

**Supplementary Figures 1-12**

**Supplementary Tables 1-10**

**Supplementary Notes 1-3**

**Supplementary Figures**

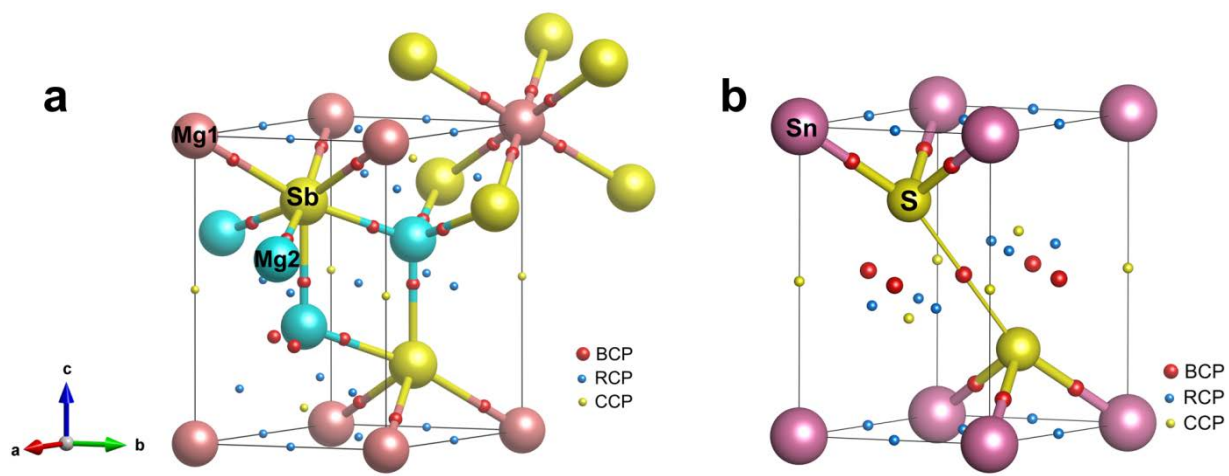

**Supplementary Figure 1.** 3D plot of critical points of electron density in (a) $Mg_3Sb_2$ and (b) $SnS_2$. BCP, RCP, and CCP denote bond critical point, ring critical point, and cage critical point, respectively.

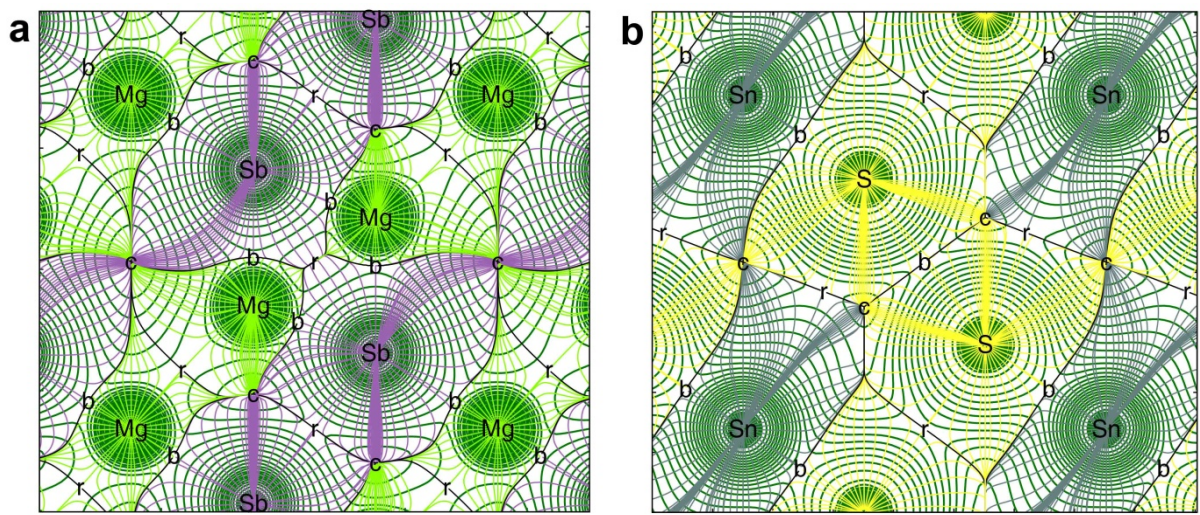

**Supplementary Figure 2.** Total electron density map with zero-flux surfaces and trajectories of the gradients on the (110) plane of (a) $Mg_3Sb_2$ and (b) $SnS_2$. r, b, and c denote ring critical point, bond critical point, and cage critical point, respectively. The olive and black curves represents the total electron density contours and zero-flux curves, respectively. The green, purple, yellow, and dark green lines denote the gradient paths.

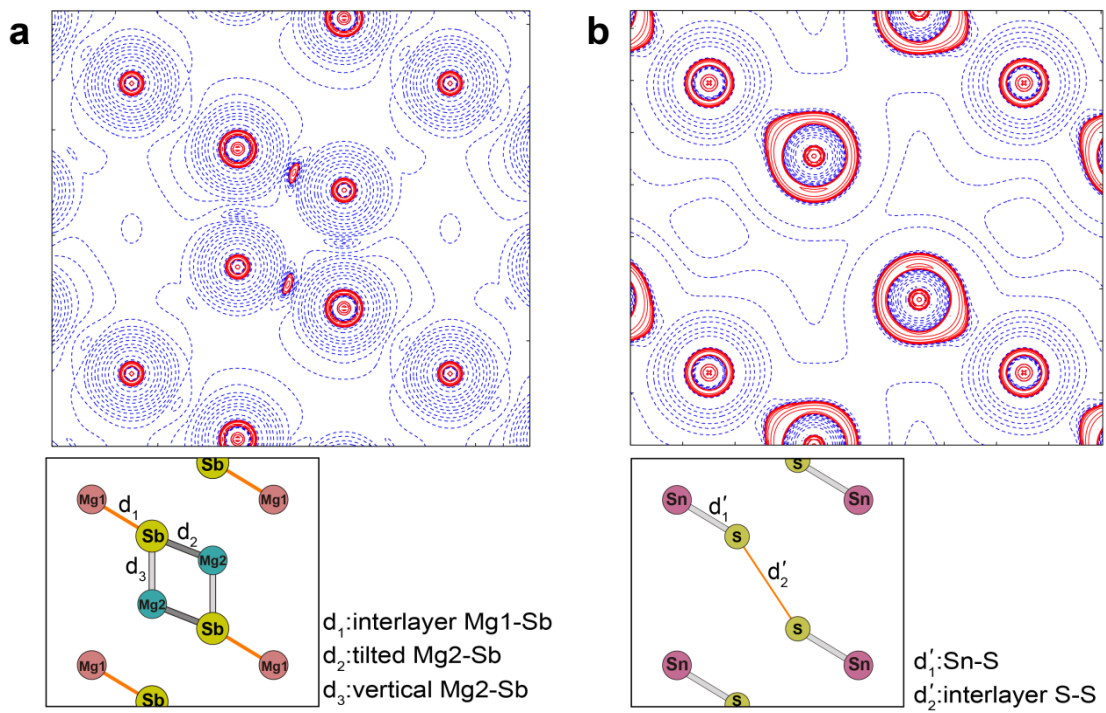

**Supplementary Figure 3.** Negative Laplacian map on (110) plane of (a) Mg$_3$Sb$_2$ and (b) SnS$_2$. Contours are drawn at $\pm 2\times 10^n$, $\pm 4\times 10^n$ and $\pm 8\times 10^n$ $e$/Å$^5$ (n = $\pm 3, \pm 2, \pm 1, 0$). Red solid and blue dotted lines represent positive and negative values, respectively. The inset shows the corresponding (110) plane.

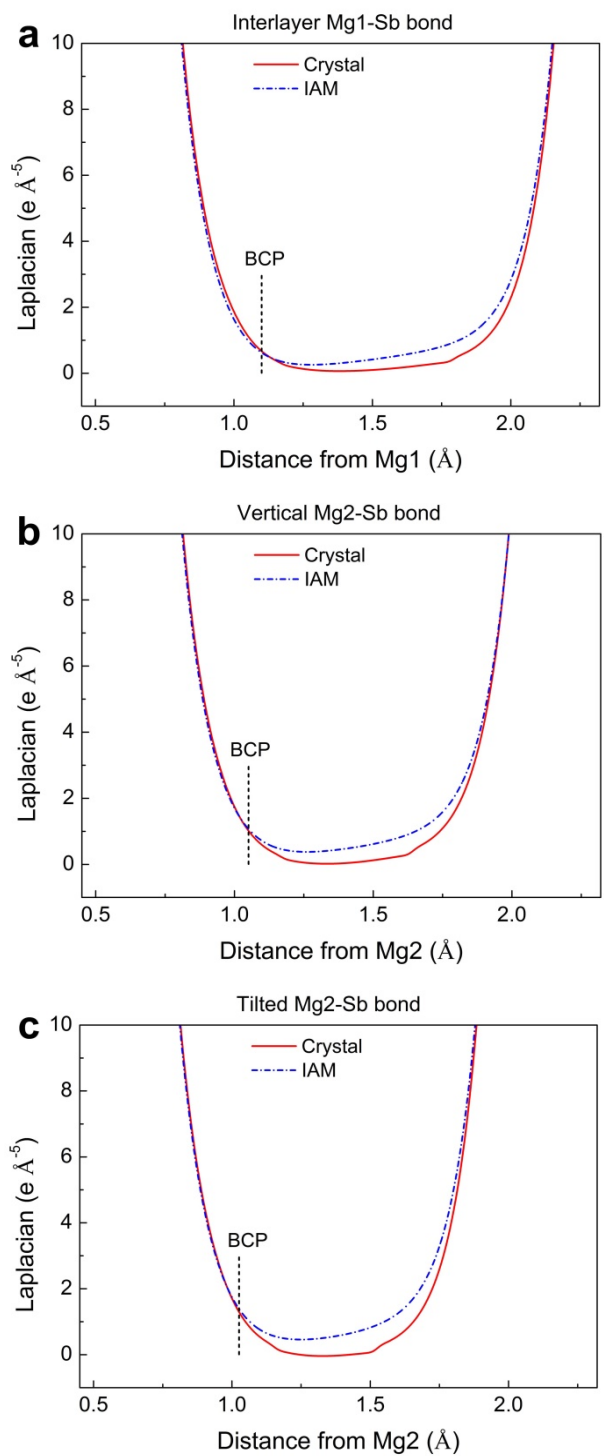

**Supplementary Figure 4.** Line plots of Lapalacian along the bond paths of (a) Mg1-Sb, (b) vertical Mg2-Sb, and (c) tilted Mg2-Sb. Red solid line represents the result based on the full electron density of $Mg_3Sb_2$, whereas the dash dot line is the result based on the electron density of Independent Atom Model (IAM). The vertical dash line marks the location of bond critical point.

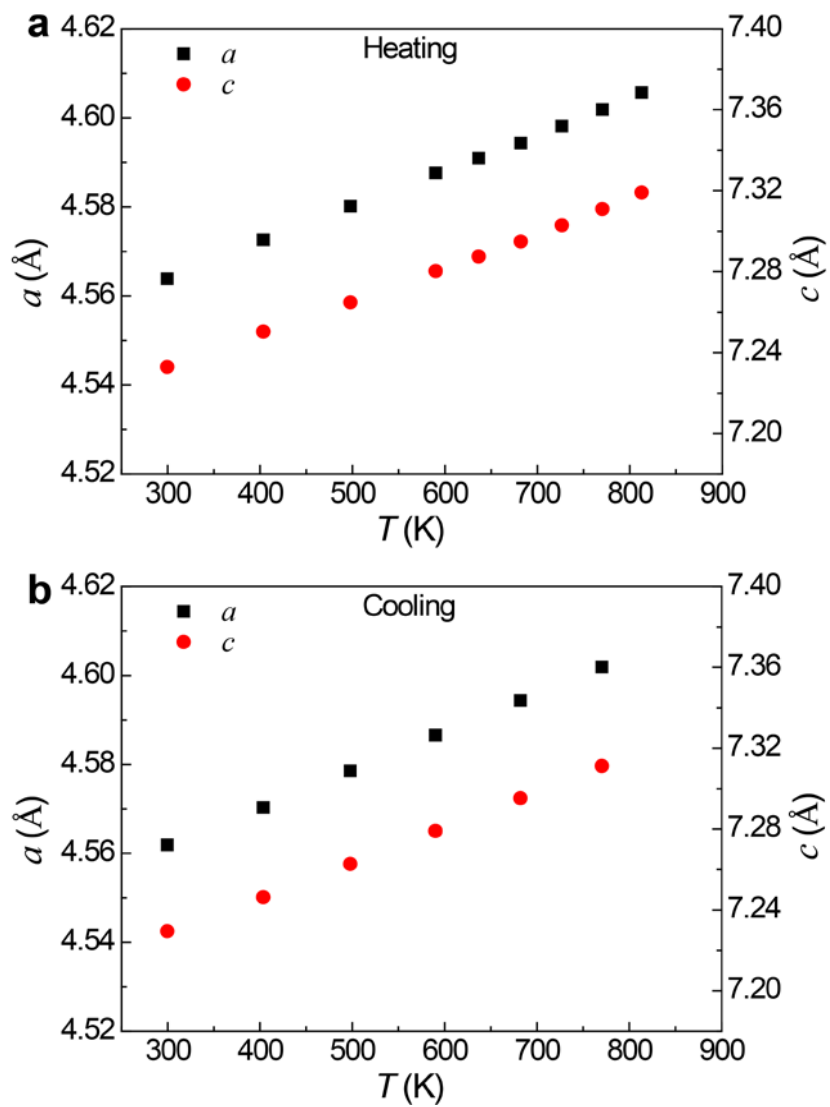

**Supplementary Figure 5.** Temperature dependence of lattice parameters of $Mg_3Sb_2$ for the temperature points of (a) 299-813 K and (b) 770-299 K. The uncertainty is smaller than the size of the symbol.

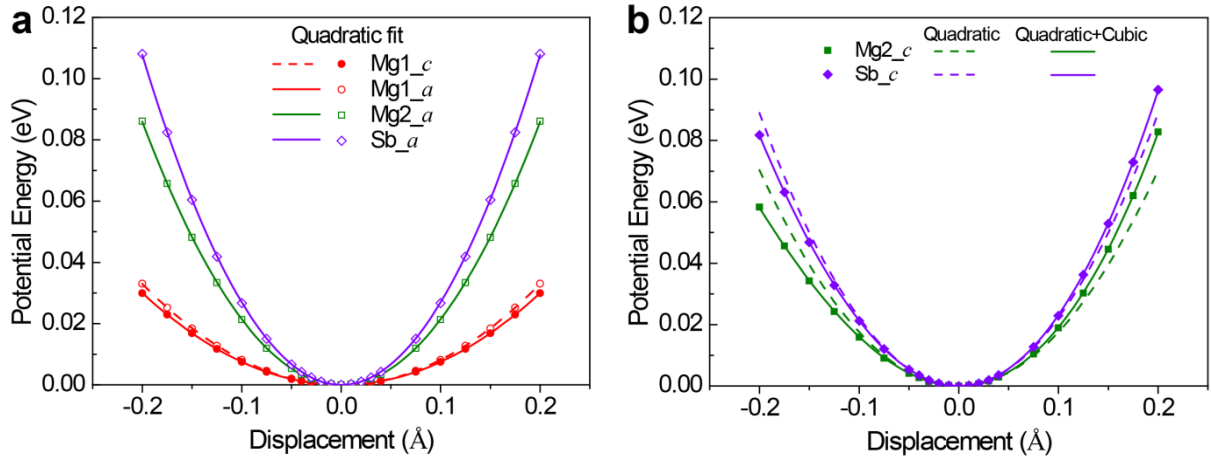

**Supplementary Figure 6.** Fitted potential energy curves for the nonequivalent atoms along the axial directions in Mg$_3$Sb$_2$. Potential wells of Mg1 along axial directions, Mg2 along the *a* direction, and Sb along the *a* direction shown in (a) are harmonic, whereas potential wells of Mg2 and Sb along the *c* direction shown in (b) are anharmonic with cubic terms.

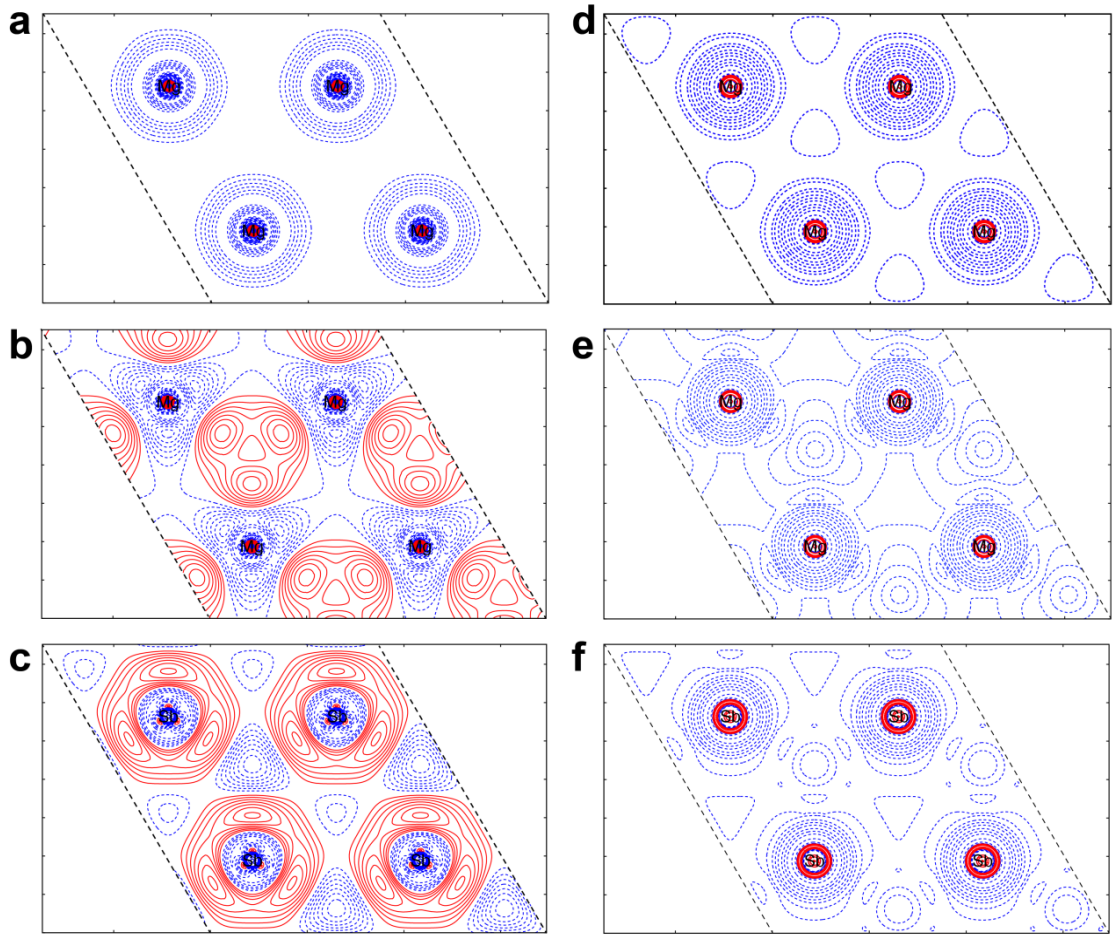

**Supplementary Figure 7. (a-c)** Static deformation maps on a-b planes containing (a) Mg1 atoms, (b) Mg2 atoms, and (c) Sb atoms of $Mg_3Sb_2$. The contour interval is 0.006 $e/Å^3$. **(d-f)** Negative Laplacian maps on a-b planes of (d) Mg1 atoms, (e) Mg2 atoms, and (f) Sb atoms of $Mg_3Sb_2$. Contours are plotted at $\pm 2\times 10^n$, $\pm 4\times 10^n$ and $\pm 8\times 10^n$ $e/Å^5$ (n = ±3, ±2, ±1, 0). Red solid and blue dotted lines represent positive and negative values, respectively.

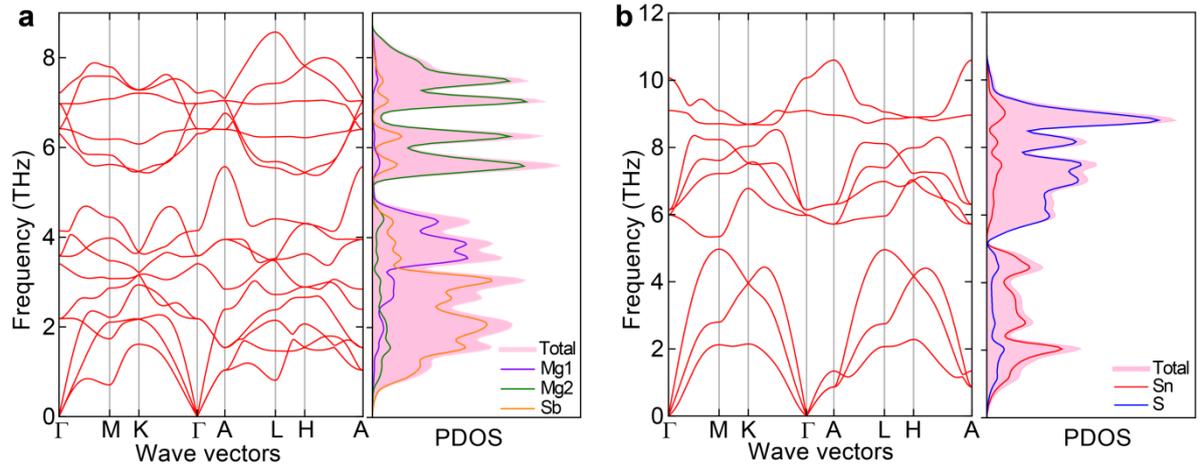

**Supplementary Figure 8.** Phonon dispersions and phonon density of states (PDOS) of (a) $Mg_3Sb_2$ and (b) $SnS_2$.

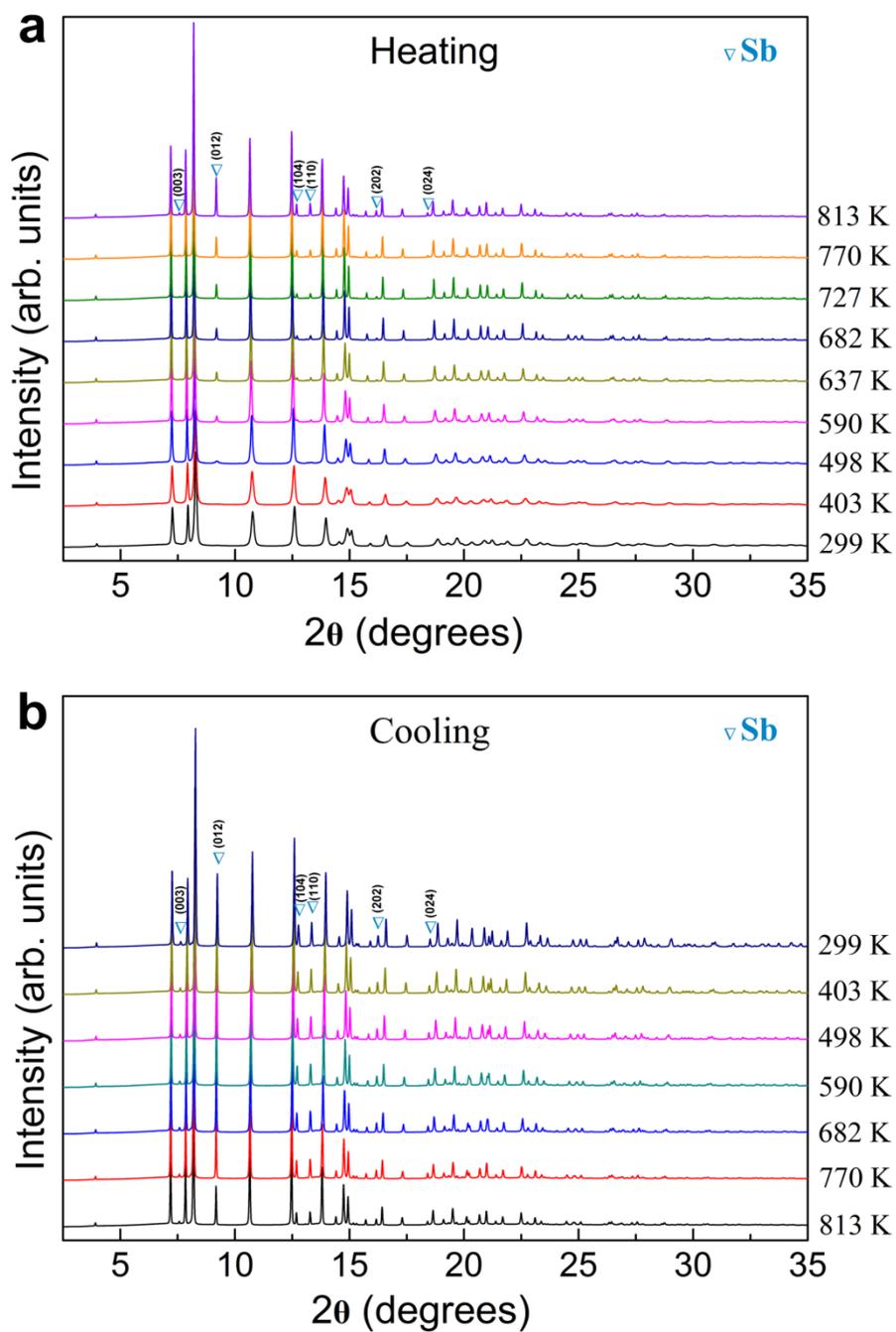

**Supplementary Figure 9.** Multi-temperature synchrotron X-ray powder diffraction patterns of $Mg_3Sb_2$ at temperatures of (a) 299-813 K (heating) and 813-299 K (cooling).

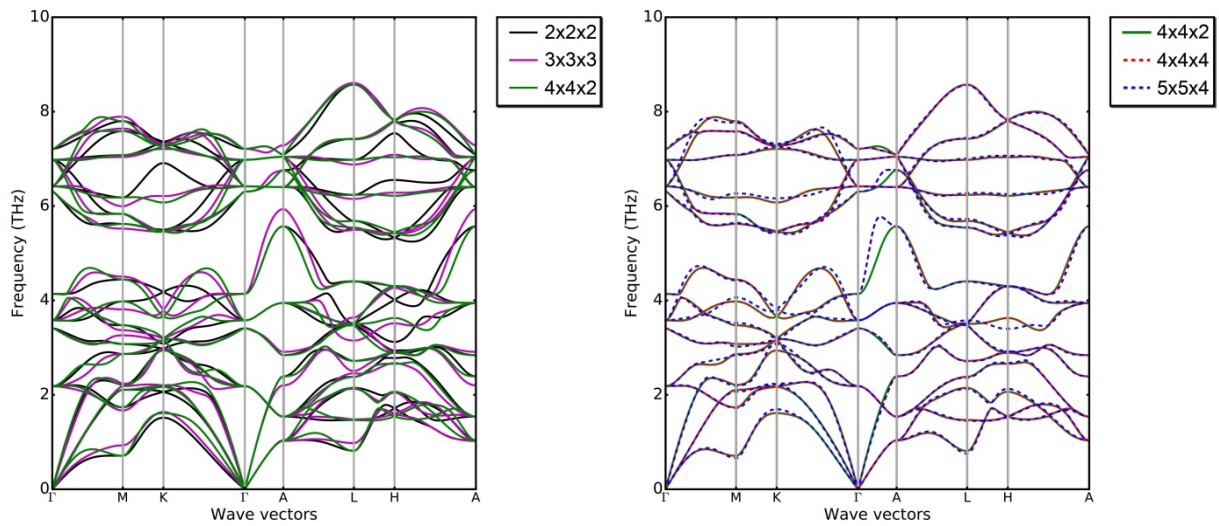

**Supplementary Figure 10.** Convergence test of phonon dispersion of $Mg_3Sb_2$ with a series of supercell sizes. It is clear that, in order to ensure the well-converged phonon frequencies of acoustic branches, at least a supercell size of 4×4×2 (160 atoms) should be used. This reveals that the reported phonon-related results in ref. 1 based on a 2×2×2 supercell are simply not converged.

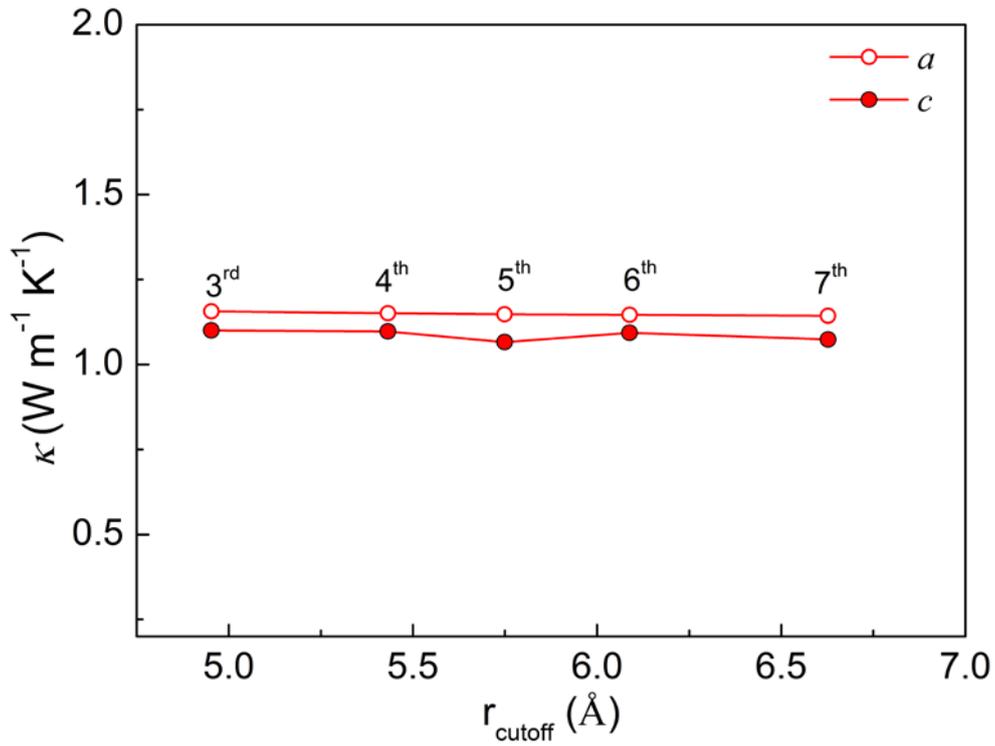

**Supplementary Figure 11.** Convergence of the lattice thermal conductivity of $Mg_3Sb_2$ with respect to the cutoff distance ($r_{cutoff}$) corresponding to the interaction range from the third to the seventh nearest neighbors for the anharmonic calculations.

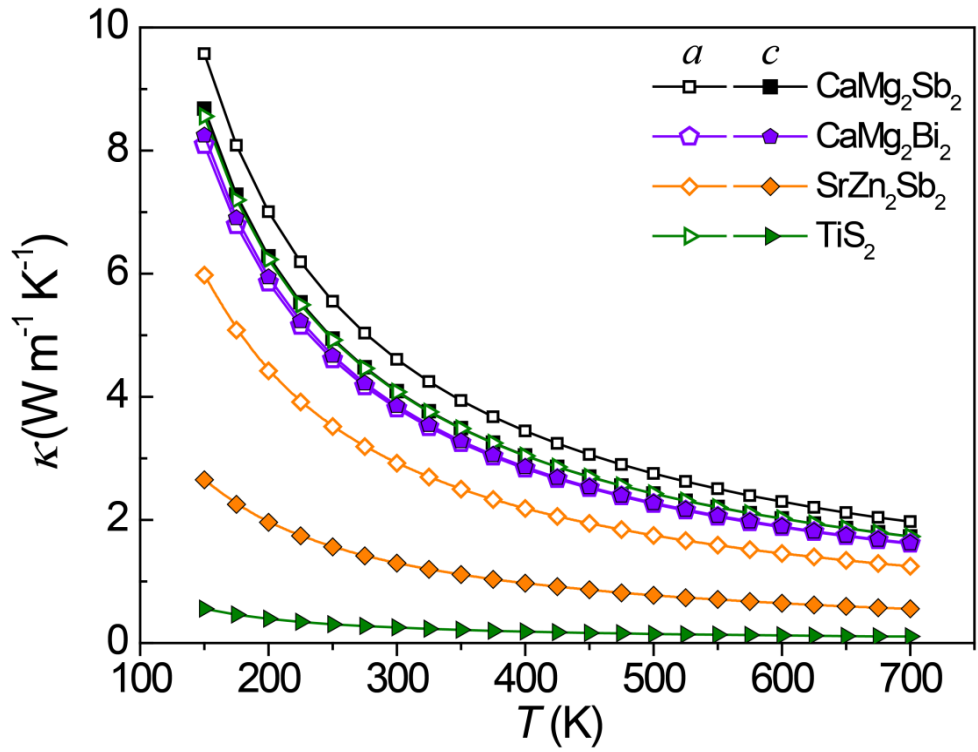

**Supplementary Figure 12.** The calculated lattice thermal conductivity of TiS$_2$, SrZn$_2$Sb$_2$, CaMg$_2$Sb$_2$, and CaMg$_2$Bi$_2$ along the *a* and *c* directions.

## Supplementary Tables

**Supplementary Table 1.** Rietveld refinement details of the synchrotron PXRD data of Mg$_3$Sb$_2$ at 770-299 K upon cooling. The *R* factors and $\chi^2$ shown here are the data from the Mg$_3$Sb$_2$ phase. $T_{actual}$ represents the actual temperature calibrated by the thermocouple.

| $T_{actual}$(K) | | 770.15 | 682.15 | 590.15 | 497.65 | 403.45 | 299.45 |
|---|---|---|---|---|---|---|---|
| No. of points | | 7727 | 7727 | 7727 | 7727 | 7727 | 7728 |
| No. of reflections | | 938 | 926 | 920 | 920 | 910 | 904 |
| No. of parameters | | 48 | 48 | 48 | 48 | 48 | 48 |
| $R_F$ (%) | | 9.50 | 7.82 | 6.91 | 4.47 | 3.38 | 2.63 |
| $R_{Bragg}$ (%) | | 4.78 | 4.69 | 4.73 | 4.88 | 4.51 | 5.60 |
| $R_p$ (%) | | 9.78 | 9.45 | 9.61 | 9.86 | 8.94 | 8.54 |
| $R_{wp}$ (%) | | 10.6 | 10.5 | 10.7 | 11.3 | 10.6 | 10.2 |
| $\chi^2$ | | 6.33 | 6.67 | 7.76 | 9.49 | 9.22 | 9.35 |
| Wt.% Mg$_3$Sb$_2$ | | 85.82(0.20) | 85.24(0.19) | 85.20(0.20) | 85.23(0.21) | 85.79(0.20) | 85.53(0.19) |
| Wt.% Sb | | 14.18(0.06) | 14.76(0.06) | 14.80(0.06) | 14.77(0.07) | 14.21(0.06) | 14.47(0.06) |
| Mg$_3$Sb$_2$ | a=b (Å) | 4.60190(5) | 4.59434(5) | 4.58658(4) | 4.57851(4) | 4.57029(4) | 4.56187(3) |
| | c (Å) | 7.31115(8) | 7.29523(8) | 7.27911(8) | 7.26271(8) | 7.24629(7) | 7.22944(6) |
| Volume (Å$^3$) | | 134.088(3) | 133.357(2) | 132.613(2) | 131.849(2) | 131.079(2) | 130.293(2) |
| $U_{iso}$ (Å$^2$) | Mg1 | 0.05806(225) | 0.05227(209) | 0.04022(191) | 0.02965(175) | 0.02312(151) | 0.01423(120) |
| | Mg2 | 0.03712(105) | 0.03175(95) | 0.02566(90) | 0.01824(82) | 0.01381(70) | 0.00907(58) |
| | Sb | 0.03404(23) | 0.02877(22) | 0.02338(19) | 0.01792(18) | 0.01374(14) | 0.00876(11) |
| Occupancy | Mg1 | 0.07591 | 0.07731 | 0.07768 | 0.07860 | 0.07842 | 0.07858 |
| | Mg2 | 0.15971 | 0.16054 | 0.16153 | 0.16288 | 0.16349 | 0.16429 |
| | Sb | 0.16838 | 0.16827 | 0.16808 | 0.16773 | 0.16763 | 0.16753 |

**Supplementary Table 2.** Lattice parameters *a* and c as a function of temperature and thermal expansion coefficients $\alpha_a$ and $\alpha_c$ at 299.45 K of $Mg_3Sb_2$. The data at 770-299 K upon cooling are used.

| Sample | *a* (Å) | $\alpha_a$ (×10⁻⁵ K⁻¹) at 299.45 K | *c* (Å) | $\alpha_c$ (×10⁻⁵ K⁻¹) at 299.45 K | $\alpha_c/\alpha_a$ at 299.45 K |
|---|---|---|---|---|---|
| $Mg_3Sb_2$ | (1st degree) *a* = 4.5360(2) + 8.56(4))×10⁻⁵*T* | 1.88(0) | (1st degree) *c* = 7.1761(8) + 17.5(1)×10⁻⁵*T* | 2.42(1) | 1.29(0) |
| | (2nd degree) *a* = 4.5369(7) + 8.2(3)×10⁻⁵*T* + 0.3(2)×10⁻⁸*T*² | 1.84(4) | (2nd degree) *c* = 7.182(1) + 15.3(4)×10⁻⁵*T* + 2.0(3)×10⁻⁸*T*² | 2.28(3) | 1.24(3) |

**Supplementary Table 3.** Group velocities of the three acoustic phonon branches along the Γ-M and Γ-A wave vector path in $Mg_3Sb_2$ and $SnS_2$. Γ-M and Γ-A correspond to the $a^*$ and $c^*$ directions, respectively.

| | $Mg_3Sb_2$ | | $SnS_2$ | |
|---|---|---|---|---|
| Wave vector path | Γ-M | Γ-A | Γ-M | Γ-A |
| TA1 group velocity (m s⁻¹) | 1969 | 1885 | 1149 | 1538 |
| TA2 group velocity (m s⁻¹) | 2021 | 1885 | 2671 | 1538 |
| LA group velocity (m s⁻¹) | 4340 | 4732 | 5297 | 2444 |

**Supplementary Table 4.** Topological properties of the bond critical points ($r_b$) in several Mg-containing compounds with $CaAl_2Si_2$-type structure.

| Bond | d (Å) | $\rho(r_b)$ (e Å$^{-3}$) | $\nabla^2\rho(r_b)$ (e Å$^{-5}$) | G (a.u.) | V (a.u.) | H (a.u.) | \|V\|/G | G/$\rho$ (a.u.) |
|---|---|---|---|---|---|---|---|---|
| \multicolumn{9}{c}{$Mg_3Bi_2$} | | | | | | | | |
| Tilted Mg2-Bi | 2.953 | 0.152 | 0.953 | 0.0118 | -0.0136 | -0.0019 | 1.160 | 0.521 |
| Vertical Mg2-Bi | 3.040 | 0.144 | 0.761 | 0.0100 | -0.0121 | -0.0021 | 1.209 | 0.468 |
| Interlayer Mg1-Bi | 3.176 | 0.107 | 0.516 | 0.0064 | -0.0075 | -0.0011 | 1.167 | 0.407 |
| \multicolumn{9}{c}{$CaMg_2Sb_2$} | | | | | | | | |
| Tilted Mg-Sb | 2.868 | 0.167 | 1.243 | 0.0146 | -0.0164 | -0.0018 | 1.120 | 0.591 |
| Vertical Mg-Sb | 2.934 | 0.156 | 1.070 | 0.0128 | -0.0144 | -0.0017 | 1.130 | 0.554 |
| Interlayer Ca-Sb | 3.278 | 0.116 | 0.781 | 0.0087 | -0.0092 | -0.0006 | 1.065 | 0.506 |
| \multicolumn{9}{c}{$CaMg_2Bi_2$} | | | | | | | | |
| Tilted Mg-Bi | 2.954 | 0.153 | 0.993 | 0.0121 | -0.0139 | -0.0018 | 1.149 | 0.533 |
| Vertical Mg-Bi | 3.009 | 0.146 | 0.860 | 0.0108 | -0.0126 | -0.0019 | 1.173 | 0.498 |
| Interlayer Ca-Bi | 3.332 | 0.109 | 0.719 | 0.0080 | -0.0084 | -0.0005 | 1.062 | 0.491 |
| \multicolumn{9}{c}{$SrMg_2Sb_2$} | | | | | | | | |
| Tilted Mg-Sb | 2.891 | 0.164 | 1.175 | 0.0140 | -0.0157 | -0.0018 | 1.127 | 0.576 |
| Vertical Mg-Sb | 2.926 | 0.155 | 1.091 | 0.0129 | -0.0145 | -0.0016 | 1.122 | 0.560 |
| Interlayer Sr-Sb | 3.407 | 0.114 | 0.726 | 0.0082 | -0.0089 | -0.0007 | 1.082 | 0.486 |
| \multicolumn{9}{c}{$YbMg_2Sb_2$} | | | | | | | | |
| Tilted Mg-Sb | 2.864 | 0.167 | 1.254 | 0.0147 | -0.0164 | -0.0017 | 1.115 | 0.595 |
| Vertical Mg-Sb | 2.939 | 0.155 | 1.052 | 0.0126 | -0.0143 | -0.0017 | 1.135 | 0.549 |
| Interlayer Yb-Sb | 3.259 | 0.148 | 0.888 | 0.0111 | -0.0130 | -0.0019 | 1.170 | 0.505 |

**Supplementary Table 5.** Topological properties of additional bond critical point found between interlayer Sb atoms. Additional bond critical point of interlayer Sb-Sb interaction is very weak and might be caused by numerical errors.

| Bond | d (Å) | $\rho(\mathbf{r}_b)$ (e Å$^{-3}$) | $\nabla^2\rho(\mathbf{r}_b)$ (e Å$^{-5}$) | G (a.u.) | V (a.u.) | H (a.u.) | $|V|/G$ | $G/\rho$ (a.u.) |
|---|---|---|---|---|---|---|---|---|
| CaZn$_2$Sb$_2$ | | | | | | | | |
| Interlayer Sb-Sb | 4.473 | 0.049 | 0.177 | 0.0020 | -0.0022 | -0.0002 | 1.082 | 0.277 |
| CaMg$_2$Sb$_2$ | | | | | | | | |
| Interlayer Sb-Sb | 4.585 | 0.048 | 0.192 | 0.0021 | -0.0022 | -0.0001 | 1.049 | 0.292 |
| CaMg$_2$Bi$_2$ | | | | | | | | |
| Interlayer Bi-Bi | 4.627 | 0.047 | 0.176 | 0.0020 | -0.0021 | -0.0001 | 1.065 | 0.279 |
| SrMg$_2$Sb$_2$ | | | | | | | | |
| Interlayer Sb-Sb | 4.720 | 0.038 | 0.144 | 0.0015 | -0.0015 | -7.5×10$^{-6}$ | 1.005 | 0.269 |
| YbMg$_2$Sb$_2$ | | | | | | | | |
| Interlayer Sb-Sb | 4.542 | 0.054 | 0.212 | 0.0024 | -0.0026 | -0.0002 | 1.078 | 0.298 |

**Supplementary Table 6.** Atomic charges $Q$ and atomic basin volumes $V$ of several Mg-containing compounds with CaAl$_2$Si$_2$-type structure.

| Atoms | $Q$ | $V$ (Å$^3$) |
|---|---|---|
| Mg$_3$Bi$_2$ | | |
| Mg1 | 1.43 | 9.79 |
| Mg2 | 1.40 | 9.81 |
| Bi | -2.11 | 57.11 |
| CaMg$_2$Sb$_2$ | | |
| Ca | 1.39 | 16.01 |
| Mg | 1.48 | 8.63 |
| Sb | -2.17 | 55.53 |
| CaMg$_2$Bi$_2$ | | |
| Ca | 1.37 | 16.65 |
| Mg | 1.42 | 9.67 |
| Bi | -2.10 | 59.14 |
| SrMg$_2$Sb$_2$ | | |
| Sr | 1.38 | 21.74 |
| Mg | 1.47 | 8.87 |
| Sb | -2.16 | 57.29 |
| YbMg$_2$Sb$_2$ | | |
| Yb | 1.34 | 19.84 |
| Mg | 1.47 | 8.57 |
| Sb | -2.14 | 52.96 |

**Supplementary Table 7.** Rietveld refinement details of the synchrotron X-ray powder diffraction data at 299-813 K upon heating. The $R$ factors and $\chi^2$ shown here are the data from the Mg$_3$Sb$_2$ phase. $T_{actual}$ represents the actual temperature calibrated by the thermocouple.

| | | 299.45 | 403.45 | 497.65 | 590.15 | 682.15 | 726.6 | 770.15 | 813.0 |
|---|---|---|---|---|---|---|---|---|---|
| $T_{actual}$ (K) | | 299.45 | 403.45 | 497.65 | 590.15 | 682.15 | 726.6 | 770.15 | 813.0 |
| No. of points | | 7727 | 7727 | 7727 | 7727 | 7727 | 7727 | 7727 | 7727 |
| No. of reflections | | 1055 | 1071 | 994 | 963 | 918 | 920 | 924 | 938 |
| No. of parameters | | 46 | 45 | 48 | 48 | 48 | 48 | 48 | 48 |
| $R_F$ (%) | | 1.37 | 1.48 | 3.31 | 4.45 | 7.21 | 9.51 | 9.84 | 10.7 |
| $R_{Bragg}$ (%) | | 5.77 | 4.34 | 8.12 | 9.99 | 5.30 | 4.91 | 5.18 | 4.24 |
| $R_p$ (%) | | 8.23 | 8.77 | 9.88 | 11.0 | 8.54 | 8.65 | 9.05 | 8.69 |
| $R_{wp}$ (%) | | 10.7 | 11.2 | 12.0 | 12.2 | 8.92 | 8.94 | 9.22 | 9.05 |
| $\chi^2$ | | 8.17 | 7.91 | 8.57 | 9.06 | 4.91 | 4.85 | 4.94 | 4.51 |
| Wt.% Mg$_3$Sb$_2$ | | 98.77(0.30) | 98.72(0.24) | 97.51(0.26) | 97.28(0.25) | 96.84(0.18) | 96.32(0.18) | 95.03(0.18) | 90.80(0.17) |
| Wt.% Sb | | 1.23(0.19) | 1.28(0.07) | 2.49(0.06) | 2.72(0.05) | 3.16(0.03) | 3.68(0.03) | 4.97(0.03) | 9.20(0.04) |
| Mg$_3$Sb$_2$ | $a=b$ (Å) | 4.56386(8) | 4.57260(8) | 4.58017(7) | 4.58760(6) | 4.59432(3) | 4.59811(3) | 4.60187(4) | 4.60569(4) |
| | $c$ (Å) | 7.23289(15) | 7.25042(15) | 7.26486(12) | 7.28032(11) | 7.29486(6) | 7.30297(6) | 7.31098(6) | 7.31910(7) |
| Volume (Å$^3$) | | 130.469(4) | 131.286(4) | 131.984(4) | 132.694(3) | 133.349(2) | 133.718(2) | 134.083(2) | 134.455(2) |
| $U_{iso}$ (Å$^2$) | Mg1 | 0.00524(121) | 0.00949(176) | 0.01672(180) | 0.02389(189) | 0.04560(152) | 0.04857(160) | 0.05315(175) | 0.05535(180) |
| | Mg2 | 0.00822(71) | 0.01344(92) | 0.02221(99) | 0.02388(98) | 0.03545(75) | 0.03699(77) | 0.03965(84) | 0.04421(91) |
| | Sb | 0.00767(18) | 0.01147(22) | 0.01782(22) | 0.02311(23) | 0.02914(15) | 0.03132(16) | 0.03402(18) | 0.03779(19) |
| Occupancy | Mg1 | 0.07336 | 0.07320 | 0.07405 | 0.07392 | 0.07780 | 0.07587 | 0.07554 | 0.07591 |
| | Mg2 | 0.16421 | 0.16440 | 0.16282 | 0.16136 | 0.16411 | 0.16240 | 0.16135 | 0.15970 |
| | Sb | 0.16720 | 0.16729 | 0.16709 | 0.16753 | 0.16729 | 0.16759 | 0.16796 | 0.16852 |

**Supplementary Table 8.** Lattice parameters *a* and c as a function of temperature and thermal expansion coefficients $\alpha_a$ and $\alpha_c$ at 299.45 K of $Mg_3Sb_2$. The data at 299-810 K upon heating are adopted.

| Sample | $a$ (Å) | $\alpha_a$ (×10$^{-5}$ K$^{-1}$) at 299.45 K | $c$ (Å) | $\alpha_c$ (×10$^{-5}$ K$^{-1}$) at 299.45 K | $\alpha_c/\alpha_a$ at 299.45 K |
|---|---|---|---|---|---|
| Mg$_3$Sb$_2$ | (1st degree) $a$ = 4.5400(4) + 8.03(6))×10$^{-5}$T | 1.76(1) | (1st degree) $c$ = 7.183(1) + 16.6(2)×10$^{-5}$T | 2.30(1) | 1.31(1) |
| | (2nd degree) $a$ = 4.540(1) + 7.9(5)×10$^{-5}$T + 0.1(4)×10$^{-8}$T$^2$ | 1.74(7) | (2nd degree) $c$ = 7.190(3) + 14(1)×10$^{-5}$T + 2.5(8)×10$^{-8}$T$^2$ | 2.14(8) | 1.23(5) |

**Supplementary Table 9.** Debye temperatures of $Mg_3Sb_2$ obtained from fitting of $U_{iso}$ with the Debye expression using the temperature points of 770-299 K upon cooling. The averaged values were estimated from fitting of averaged $U_{iso} = 1/5\ U_{iso}(Mg1) + 2/5\ U_{iso}(Mg2) + 2/5\ U_{iso}(Sb)$. The values of Debye temperature obtained by fitting using the Debye expression are compared with the value reported in the literature[2].

| Compound | Atoms | Debye temperature $\Theta_D$ (K) | |
|---|---|---|---|
| | | Debye expression (This work) | Calculated from elastic constants[2] |
| $Mg_3Sb_2$ | Mg1 | 248(6) | - |
| | Mg2 | 311(7) | - |
| | Sb | 149(2) | - |
| | Average | 187(4) | 223 |

**Supplementary Table 10.** The average Grüneisen parameter at 300 K along the *a* and *c* directions obtained using Equation 3.

| Compounds | $\tilde{\gamma}_a$ | $\tilde{\gamma}_c$ | $\tilde{\gamma}_c / \tilde{\gamma}_a$ |
|---|---|---|---|
| $Mg_3Sb_2$ | 2.1 | 2.6 | 1.2 |
| $SnS_2$ | 2.0 | 3.7 | 1.9 |

## Supplementary Notes

**Supplementary Note 1. Topological analysis of electron density**

The quantitative analysis of chemical bonding in this work is based on Bader's quantum theory of atoms in molecules (QTAIM)[3]. QTAIM is based on the analysis of critical points (CPs) of the electron density, which are defined as the points satisfying $\nabla\rho = 0$. In general, CPs are classified according to the *rank*, i.e., the number of nonzero eigenvalues of the Hessian matrix, and the *signature*, which is the sum of the signs of the Hessian eigenvalues. These two characteristics are then used to label the CPs (*rank*, *signature*)[3,4]. Accordingly, there are four types of rank 3 CPs: (3, –3), i.e., maxima or nuclear CPs (NCP or n); (3, –1), first-order saddle or bond (BCP or b); (3, +1), second-order saddle or ring (RCP or r); and (3, +3), minima or cage (CCP or c). The illustration of critical points in $Mg_3Sb_2$ is shown in Supplementary Figs. 1 and 2. An atom in a molecule or crystal is defined as the space with a density maximum surrounded by a zero-flux gradient surface $S$:[3,4]

$$\nabla\rho(\mathbf{r})\cdot\mathbf{n}(\mathbf{r}) = 0 \quad \forall \mathbf{r} \in S(\mathbf{r}_s), \tag{1}$$

where $\mathbf{n}(\mathbf{r})$ is the a unit vector perpendicular to the surface $S$ at $\mathbf{r}$. An atom is described as the union of an attractor and its basin. The atomic property of an atom is then calculated by the integration within its atomic basin.

**Supplementary Note 2. Synchrotron PXRD patterns and Rietveld refinement**

From multi-temperature synchrotron PXRD patterns of $Mg_3Sb_2$ shown in Supplementary Fig. 9, we can see that the Sb phase appears as the temperature increases. To keep consistence, the secondary phase Sb is included in the refinements of all temperature points. Although there is small difference between the thermal expansion coefficients calculated by the data upon heating and cooling, the anisotropic ratio of thermal expansion coefficient upon heating and cooling is nearly the same (see Supplementary Tables 8 and 2). As shown in Supplementary Table 7, the amount of Sb phase is gradually increasing with increasing temperature from 299 K to 810 K, whereas the amount of Sb phase becomes very stable upon cooling from 770 K to 299 K (see Supplementary Table 1). To avoid the possible influence from the increasing Sb phase upon heating, we thereby use the cooling data at 770-299 K with the stable amount of Sb in the main text. The underlying mechanism of the appearing Sb phase upon heating in Mg$_3$Sb$_2$ powder will be discussed in our future work.

The Debye temperature was extracted by fitting the isotropic atomic displacement parameters $U_{\text{iso}}$ based on a Debye model:[5-7]

$$U_{\text{iso}} = \frac{3\hbar^2 T}{mk_B \Theta_D^2}\left[\frac{T}{\Theta_D}\int_0^{\frac{\Theta_D}{T}}\frac{x}{e^x-1}dx + \frac{\Theta_D}{T}\right] + d^2, \quad (2)$$

where $\hbar$ is the reduced Planck constant, $T$ the absolute temperature, $m$ is the mass of the atom, $k_B$ is the Boltzmann constant, $\Theta_D$ is the Debye temperature, and $d$ is a disorder parameter[8]. The Debye temperatures of Mg$_3$Sb$_2$ obtained from fitting of $U_{\text{iso}}$ at 770-299 K (upon cooling) with the Debye expression are shown in Supplementary Table 2. The average Debye temperature obtained by fitting using the Debye expression is slightly lower than the reported theoretical value[2] calculated from elastic constants of Mg$_3$Sb$_2$.

**Supplementary Note 3. Theoretical calculation**

The Grüneisen parameters for the acoustic modes might be negative, which will lead to the cancellation between the acoustic and optical modes. In addition to the average method used in the main text, the average Grüneisen parameter is calculated using the sum over the squared Grüneisen parameter:[9]

$$\tilde{\gamma}^2 = \frac{\sum_{\mathbf{q},i}[\gamma(\mathbf{q},i)]^2 C_V(\mathbf{q},i)}{\sum_{\mathbf{q},i} C_V(\mathbf{q},i)}, \quad (3)$$

where $\gamma(\mathbf{q},i)$ and $C_V(\mathbf{q},i)$ are the mode Grüneisen parameter and mode heat capacity for the phonon branch $i$ at wave vector $\mathbf{q}$, respectively. The average Grüneisen parameters along the axial directions were calculated by sum over all phonon modes along the corresponding directions and the result is shown in Supplementary Table 10.

The lattice thermal conductivity of TiS$_2$ was calculated by ShengBTE code[10] based on a full iterative solution to the Boltzmann transport equation for phonons. The second-order and third-order interatomic force constants were computed in the 4×4×2 supercells. The displacement amplitude of 0.09 Å was adopted for harmonic force constants calculations to ensure the well-converged properties[11]. Van der Waals functional optB86b-vdW[12] in VASP[13] code was used for all calculations with an energy convergence criterion of $10^{-6}$ eV and a plane wave energy cutoff of 600 eV.